# Response of Superheated Droplet Detectors of the PICASSO Dark Matter Search Experiment

M. Barnabé-Heider, M. Di Marco[§], P. Doane, M.-H. Genest, R. Gornea, R. Guénette,
C. Leroy[*], L. Lessard, J.-P. Martin, U. Wichoski and V. Zacek
*Département de physique, Université de Montréal, Montréal, H3C 3J7, Canada*

K. Clark, C.B. Krauss and A.J. Noble
*Department of physics, Queens University, Kingston, K7L 3NG, Canada*

E. Behnke, W. Feighery[†], I. Levine and C. Muthusi[†]
*Department of physics & astronomy, [†] Department of Chemistry,
Indiana University South Bend, South Bend, Indiana, 46634, USA*

S. Kanagalingam and R. Noulty
*Bubble Technology Industries, Chalk River K0J 1J0, Canada*

Abstract

We present results of systematic studies of the radiation response of superheated liquid droplet detectors, which are used in the PICASSO dark matter search experiment. This detection technique is based on the phase transitions of superheated liquid Freon droplets dispersed and trapped in a polymerized gel. Phase transitions can be induced by nuclear recoils following particle interactions and, in particular, interactions with Weakly Interacting Massive Particles (WIMPs). These detectors are threshold devices since a minimal energy deposition is necessary to induce a phase transition and their sensitivity to various types of radiation depends strongly on the operating temperature and pressure. The sensitivity to neutrons, α-particles and γ-rays was determined as a function of these operating parameters and the results are compared with simulations. In particular, we present a complete characterization of the response of detector modules already in use for a dark matter search at the SNO site to detect WIMPs and discuss possible background sources.

---

[*] Corresponding author : Bureau LRL 235, Département de physique, Université de Montréal, C.P. 6128, succ. centre-ville, Montréal (Qc), H3C 3J7, Canada. Tel.: 1-514-343-6722.
*E-mail address* : leroy@lps.umontreal.ca.

[§] Current address : Department of physics, Queens University, Kingston, K7L 3NG, Canada.

# 1. Introduction

The PICASSO (Project In CAnada to Search for Supersymmetric Objects) collaboration uses superheated droplet detectors to search for dark matter candidate particles, and, in particular, for Weakly Interacting Massive Particles (WIMPs). A prototype experiment based on this technique has been conducted at the underground laboratory of the Sudbury Neutrino Observatory and first physics results have been obtained [1]. We present in the following a description of the response of these devices to WIMP-induced nuclear recoils and to possible background sources.

Superheated droplet detectors are based on the bubble chamber principle [2]. In this technique, the detector medium is a metastable, superheated liquid, and a phase transition is triggered by heat spikes produced by the energy deposited along the track of a traversing charged particle. The application of this technique for dark matter searches was first proposed in [3,4] and first results were reported by the PICASSO and SIMPLE groups [5,6]. In its technical realization, a PICASSO detector consists of an emulsion of microscopic metastable superheated droplets of fluorinated halocarbons, such as $C_3F_8$ or $C_4F_{10}$, dispersed in a water-based cross linked polymer. The polymer provides a nearly ideal interface surrounding each droplet and despite its large overall surface area spontaneous boiling at the droplet interface is negligible. This contrasts with the normal bubble chamber operation, where boiling occurs at impurities and defects at the container wall which limits the sensitive time of the metastable liquid. After expansion the bubbles trapped in the polymer can be re-compressed back to droplets. Therefore droplet detectors are, apart from the recompression periods, 100% active and can easily be calibrated with high count rates with radioactive sources.

Due to the energy deposited by incoming radiation, a droplet is entirely converted during the liquid-to-vapour phase transition into a bubble trapped in the polymer at the site of its formation. This technique was developed by Ing et al. [7], driven by the need of the nuclear industry for safe and reusable personal neutron dosimeters and devices of this kind are commercialized by Bubble Technology Industries [8] under the name "bubble detectors". For small volumes (and low gas loading), as is typical in dosimetry applications, the bubbles accumulated for a period of time can be counted visually and the number of bubbles recorded is directly related to the radiation dose received. More details and descriptions of applications can be found in [7, 9]. Similar devices were developed by Apfel [10] with the difference that the droplets are dispersed and maintained suspended in a viscous gel. Following the burst of a droplet, the resulting bubble migrates to the surface of the gel and the amount of gas collected is a measure of the received dose.

The liquid-to-vapour phase transition can be induced by nuclear recoils (through interactions with neutrons or other particles), or by direct energy deposition (as is the case for gamma, beta, and alpha particles). Bubble detectors are threshold detectors, since a minimal energy deposition is needed in order to induce a phase transition. Their sensitivity to various types of radiation depends strongly on the operating temperature and pressure. The liquid-to-vapour transition is explosive in nature and is accompanied by an acoustic shock wave which can be detected with piezoelectric transducers.

Over the last several years, BTI and the PICASSO group have been collaborating in making bubble detectors using detector fabrication characteristics that are appropriate for the application to dark matter searches. To distinguish these detectors from the conventional commercial bubble detectors, these detectors have been named "Special Bubble Detectors (SBD)". More recently, detectors of large volume (up to 4.5 litres) of the types shown in Fig.1 have been developed by the PICASSO group. A set of 1 litre detectors was installed at the Sudbury Neutrino Observatory (SNO) to perform a direct search for cold dark matter (CDM) particles, in particular neutralinos, (weakly interacting particles which are predicted by minimal supersymmetric models [11]). First results of these measurements were reported in [1].

Section 2 presents the principle of operation of SBDs. The detection principle can be understood within the framework of the theory of Seitz [12], which describes the process of vaporization of superheated liquids. The fabrication of SBDs is reviewed in section 3. The response of SBD's to radiation depends on the droplet size and the active mass in the detector. Measurements of the distributions of droplet diameters for various detector samples, and several techniques to determine the amount of active mass (loading) in the form of droplets are presented in section 4. Section 5 reviews signal measurements by piezo-electric sensors of the acoustic shock wave that accompanies the explosive liquid-to-vapor transition. The sensitivity of SBDs to nuclear recoils is discussed in section 6. From purely kinematical considerations, nuclear recoil thresholds in SBDs can be obtained in the same range for neutrons of low energy (from 10 keV up to a few MeV) and massive neutralinos (from 10 $GeVc^{-2}$ up to 1$TeVc^{-2}$) at velocities which are typical for dark matter particles in our galactic halo. In this operating range the detectors are insensitive to minimally ionizing particles and γ radiation. Measurements of the SBD response to neutrons are presented and compared to simulations. The simulations use a consistent set of variables, which parameterize the underlying model of recoil energy threshold and energy deposition, following the theory of Seitz. The problem of alpha contamination in the detector is addressed in section 7. Heavy salt and other ingredients, mixed in the detector matrix during detector production, contain contaminants, which are α-emitters, such as U/Th and their daughter nuclei. The α background is one of the main backgrounds at normal temperatures of SBD operation (since other potential backgrounds, such as γ-rays and minimally ionizing particles, contribute to the detector signal only when the detectors are operated at much higher temperatures.) Therefore the response of SBDs to α-particles has to be well understood and measurements and simulations of the SBD response to α-particles are presented. Gamma response measurements are presented in section 8. A discussion of the suitability of droplet detectors for dark matter searches is given in section 9, where the neutralino response is discussed as a function of temperature. The conclusions are drawn in section 10.

## 2. Detector principle

The response of SBDs to incoming particles is entirely determined by the particle's linear energy transfer (LET), the operating temperature, the external pressure $p_o$ applied to the detector, and the intrinsic properties of the active liquid: *i)* its surface tension $\sigma(T)$; *ii)* its critical temperature $T_c$, i.e. the temperature above which the liquid phase can no longer exist; *iii)* its boiling temperature $T_b$, i.e. the temperature above which a fluid becomes either a vapour or a superheated liquid; *iv)* its vapour pressure at a given temperature $p_v(T)$. The detector operation can be understood within the framework of Seitz's theory [12] of bubble chamber operation. In this model, heat spikes, which are produced by the energy deposited by a charged particle on its track, trigger the formation of vapour bubbles in the superheated liquid.

For a given temperature in the range between $T_b$ and $T_c$, the liquid will be thermodynamically stable if the external pressure $p_o$ is larger than the vapour pressure $p_v$ and the detector will not be sensitive. If the external pressure is decreased below $p_v(T)$, droplets become superheated with a tendency to undergo a transition from the liquid (high potential energy) to the gaseous phase (lower potential energy); the difference $\delta p = p_v - p_o$ represents the degree of superheat.  However, if left undisturbed, the liquid remains in its metastable state since it must overcome a potential barrier to make the transition from the liquid to the gas phase. This can occur  if the droplet receives an extra amount of energy delivered by a heat spike due to the energy deposited by incoming particles. The potential barrier is given by Gibbs' equation [13]

$$E_c = \frac{16\pi}{3} \frac{\sigma(T)^3}{(p_v(T) - p_0)^2} \qquad (1)$$

where the surface tension at the liquid-vapour interface at temperature $T$ is given by $\sigma(T) = \sigma_0(T_c-T)/(T_c-T_0)$ and $\sigma_0$ is the surface tension at a reference temperature $T_0$, usually the boiling temperature $T_b$.

Bubble formation will occur when the deposited energy, $E_{dep}$, exceeds the threshold energy $E_{th} > E_c$ within a distance $L_c \approx aR_c$, where the so called critical radius $R_c$ is given by

$$R_c = \frac{2\,\sigma(T)}{(p_v - p_0)}\,. \qquad (2)$$

A value of $a \approx 2$ is suggested in Ref. [14], but higher values, up to $a \approx 13$, are given in Ref. [15]. An even higher value of $a \approx 18$ is found from our simulations of the α-particle and neutron responses (see Sections 6 and 7). If $dE/dx$ is the mean energy deposited per unit distance, or the linear energy transfer (LET) in the liquid, then the energy deposited along $L_c$ is $E_{dep} \approx dE/dx \cdot L_c$. Therefore, the condition to trigger a liquid-to-vapour transition becomes $E_{dep} \geq E_{th}$. However it is not the total deposited energy that will trigger a liquid-to-vapour transition, but rather the fraction of this energy transformed into heat. Therefore the actual minimum or threshold energy, $E_{th}$, for radiation or recoil nuclei detection is related to $E_c$ by an efficiency factor, $\eta = E_c/E_{th}$, where $\eta$ can range from 2 to 6% [10,16].

For a deposited energy smaller than $E_{th}$, no bubble formation occurs, but the more the deposited energy exceeds $E_{th}$, the more probable a possible explosive liquid-to-vapour transition becomes. Therefore the transition probability is not a step function, but exhibits a gradual increase as a function of deposited energy, or equivalently as a function of excess temperature above the threshold temperature until full efficiency is achieved. This interpretation follows from measurements of the thermal neutron response of detectors using various active Chlorine-loaded gases as a function of temperature [8]. These measurements confirm that the superheated liquids under investigation become increasingly sensitive to the thermal neutron-induced reaction $^{35}Cl(n,p)^{35}S$ as the temperature increases above a certain threshold (the detected recoiling $^{35}S$ nucleus deposits 17 keV energy and the proton leaves undetected with 596 keV).

The progressive behaviour of the count rate around threshold from [9] can be well described by a (sigmoid) threshold function, where the probability $P(E_{dep}, E_{th})$, that an energy deposition $E_{dep}$ in a liquid with an energy threshold $E_{th}$ will generate an explosive droplet-bubble transition is given by

$$P(E_{dep}, E_{th}(T)) = 1 - \exp\left[\frac{-\alpha(E_{dep} - E_{th}(T))}{E_{th}(T)}\right] \quad (3)$$

where $\alpha$ is a free parameter obtained from fits to the data.

Some liquids such as $C_3F_8$, $C_4F_{10}$, $CF_3Br$, $CCl_2F_2$ are especially attractive for particle detection, since they are superheated at room temperature and ambient pressure in the absence of heterogeneous nucleation. For a combination of two gases, such as $C_3F_8$ and $C_4F_{10}$, $T_b$ and $T_c$ can be adjusted by varying the relative gas concentrations in the mixture. In our studies two types of detectors have been used: SBD-100 and SBD-1000. The SBD-100 detectors, loaded with a mixture of fluorocarbons (50% $C_3F_8$, 50% $C_4F_{10}$) [17], are sensitive to neutrons with energies above 100 keV at atmospheric pressure and room temperature (~20°C). The SBD-1000 detectors, loaded with 100% $C_4F_{10}$ gas [17], are sensitive to neutrons above 1000 keV at atmospheric pressure and room temperature (~20°C). For instance $T_b$ = 248.7 K, $T_c$ = 365.7 K for a SBD-100 detector, whereas $T_b$ = 270.7 K, $T_c$ = 386.3 K for a SBD-1000 detector.

In order to compare the responses of different liquids d'Errico introduced in [9] the reduced superheat variable $s$ defined as

$$s = \frac{(T-T_b)}{(T_c-T_b)} \quad (4)$$

By definition, one has $0 \leq s \leq 1$. For any gas, $s = 0$ for $p_0 = p_v$ or $T = T_b$ which represents the lower limit of superheat, while $s = 1$ for $T = T_c$ which represents the upper limit of the superheated state. This limit of metastability of a superheated liquid is described in [18] and references therein. It was observed that for organic liquids the fluid phase ceases to exist at a temperature which is about 90% of the tabulated critical temperature $T_c$ (in Kelvin). Therefore, an effective value of the reduced superheat is used in Eq. (4) with $T_c$ replaced by $0.9\ T_c$. It will be shown in Sections 4 and 5 that this modified definition of superheat also yields a better representation of our measured alpha and neutron responses.

Since $E_{th}$ depends on the linear energy transfer of the incoming radiation, as well as the operating temperature and the pressure applied to the liquid, the detector can be set into a regime where it responds mainly to nuclear recoils. This allows discrimination against background radiations such as minimally ionizing particles and gamma rays (Fig. 2). This feature allows the operation of SBD-100 and SBD-1000 detectors at ambient pressure and close to room temperature in a mode where they are exclusively sensitive to nuclear recoils such as those following WIMP interactions. This yields a powerful background discrimination. In contrast to alternative techniques, this technique is insensitive to β and γ radiation and to cosmic ray muons, and avoids the need for cryogenics. An additional advantage of this technique is the use of an active material of relatively low cost which is readily available in large quantities. Detectors of enlarged volume beyond 1 litre with increased active mass up to 4.5 litres can be fabricated and organized in arrays of modules to obtain a large-mass system in the 10 to 100 kg range needed for the search of WIMPs predicted [19] in the framework of the minimal supersymmetric model. An array of six 1 litre detector modules was used in the dark matter search at the SNO site to detect WIMPs and possible background sources [1]. Moreover, phase transition events in these detectors can be recorded with piezoelectric transducers of moderate cost. These transducers, connected to an appropriate data acquisition system, allow efficient event localization in large volumes.

3. Detector fabrication

The PICASSO detector is based on a liquid Freon droplet emulsion dispersed in a polymerized gel and hence consists of two basic components: i) the active component which is a metastable liquid at the detector's operating temperature and pressure; and ii) the support matrix which is a tri-dimensional polymerized aqueous gel. The 1 litre detector (Fig. 1) containers used for the PICASSO dark matter search are made of natural polypropylene (HIMONT SV258) selected for radio-purity and tested for radon emanation (15 ± 9 atoms/d of Rn) [20,22] and diffusion (suppression factor of $10^5$)[21,22]. The purification of the detector components and the detector fabrication itself are performed in a clean-room environment.

The first step of the fabrication process is to purify all the chemical ingredients that enter in the preparation of the gel, including the water. The ingredient that contributes most to the radioactive contamination is CsCl, a heavy salt that is dissolved in the water of the aqueous gel to equalize the densities of the droplets and the solution and which is necessary to obtain a uniform droplet dispersion. This salt is naturally contaminated with actinides and α-emitting daughters of Th and U. The purification process consists of dissolving the salt in water (10% solution), adding Hydrous Titanium Oxide (HTiO) [22,23] (a precipitate that attracts contaminants) and filtering the final solution using mechanical filters with meshes with pore sizes as fine as 0.1 μm. The purification achieved thus far was assayed at the level of $10^{-10}$ gU/g. All the other gel ingredients are purified using the same method. The active Freon liquid itself is distilled before use in order to remove radioactive contamination. A detailed description is found in [22].

In the second step, the ultra pure aqueous solution containing the chemical agents necessary for the creation of a tri-dimensional polymer – the gel matrix – is prepared, using the purified ingredients. This solution is then degassed under vacuum, poured into the detector container and cooled to a temperature well below the boiling temperature $T_b$ of the active liquid. This ensures that the active component will remain liquid when it is added to the monomer.

In the third step, the active liquid is poured into the detector container. The container is then hermetically sealed and transferred to a water basin and allowed to thaw. When the solution, at this point still a monomer, becomes liquid, a vortex of determined speed is induced in the container for about one minute in an off-axis centrifuge. As soon as the vortex forms, droplets of the active liquid start to be dispersed.

The equality between the monomer and the active liquid densities (ρ~1.3-1.6 g/cm$^3$) along with the viscosity of the solution maintain the droplets in suspension, until in the final step, the monomer undergoes polymerization. Since the polymerization reaction is exothermic and the amount of heat produced is much larger than the thermal losses through the container wall, the temperature of the detector increases rapidly [22].

For large detectors, the heat increase can be large enough to bring the average temperature of the detector beyond the critical temperature, $T_c$. To avoid this, the detector is installed in a pressure vessel where the solution is kept pressurized at about 40 bars for a period of at least 6 hours to allow the consolidation of the polymerized gel structure.

During the whole fabrication process, precautions have to be taken in order to keep metals from coming into contact with the corrosive CsCl solution, which would introduce impurities and radioactive contamination. The effect of purification on the background count rate is shown in Fig. 3 for the same type of 1 litre detectors. The count rate decreased by three orders of magnitude from the time when no purification was performed before fabrication until the time when all ingredients were purified.

4. Droplet size and active mass

The detector sensitivity to radiation depends on the droplet size and the active mass. The droplet diameter distribution is a function of the relative shear velocity between the Freon and the aqueous monomer and the Freon viscosity during vortexing. The distributions were measured by direct observation under a microscope for several 1 litre detectors. To obtain the distributions, samples were taken from the detector gel at various locations within the detector. The samples were then scanned with a previously calibrated transmission microscope coupled to a digital camera [24]. For the study of the 1 litre detectors, 6 mm$^3$ of gel were analysed, containing roughly 8000 droplets, and the relative amount of Freon mass (loading), in g/g, was found to be 0.36±0.04%. The droplet diameter distributions, for the 1 litre detectors peak around 5 µm and deviate (Fig.4) from the distribution found in commercial 10 ml detectors used in early stages of the project [3,4]. The droplet size distribution was found to be independent of the sampling depth in the detector.

Four techniques have been used to determine the amount of active mass (loading) in the form of droplets in our detectors. The results from all these different approaches agree.

*i)* One way to determine the active mass in the form of droplets in a detector, is to measure the response of the detector to a neutron beam of well-defined energy within a given temperature range. Knowing the neutron flux, the interaction cross section, the bubble production probability at a given energy and temperature and the chemical composition of the liquid droplets, one can infer from the measured count rates the fraction of active mass in a detector module. This fraction of active mass is given by

$$f = \frac{N \cdot A}{n \cdot N_A \cdot M \sum_i \varepsilon_i \, \sigma_i \, N_i} \qquad (5)$$

where $N$ is the number of recorded events, $M$ is the detector mass, $n$ is the neutron flux integrated over the measuring period, $N_A$ is Avogadro's number, $A$ is the mass number of the gas molecule, $\sigma_i$ is the neutron cross section, $N_i$ is the number of nuclei and $\varepsilon_i$ is the efficiency for species $i$ as given by Eq. (11) in section 4. In the case of a detector of 12 g total mass with a sensitivity $S = 7 \times 10^{-3}$ counts $n^{-1} cm^{-2}$, the fraction of active mass is found to be $f = 0.70 \pm 0.06\%$. The active masses of two SBD's of volume 10 ml were measured using this method.

*ii)* A visual method has been used to determine the active mass of four 1 litre detectors. The spatial distribution of the droplets in the gel and large samples of droplet diameters have been measured using the same visual method as for the droplet size measurement. The error on the active mass depends on the volume of the analyzed gel and the uncertainty in the microscope calibration (5%).

*iii)* A direct measurement of the active mass was performed by weighing the components of a 1 litre detector during the fabrication stage prior the addition of the active liquid with a scale of 0.1g precision. The total weight was subtracted from the weight of the detector after polymerisation, including the active liquid trapped in the gel matrix, yielding the active mass of the detector. The uncertainty of this method is around 2% and determined by the scale error and some condensation of water vapour from the surrounding air on the outside of the cold detector walls during the fabrication. Once the active mass is determined, it is related to the response of the detector to neutrons measured at a fixed distance (1m) from a calibrated AcBe source. The resulting count rate, expressed in counts $g^{-1}n^{-1}cm^{-2}$ serves as a calibration standard. These neutron calibrations have been carried out at several specific temperatures (around 25°C for SBD-100 which is equivalent to 45°C for SBD-1000).

*iv)* Monte Carlo simulations were used to relate the active mass to the response of 1 litre detectors, placed 1m away from a calibrated AcBe neutron source (section 5). The experimental detection efficiency of the detector, expressed in counts $n^{-1}cm^{-2}$, was divided by the efficiency found by the simulations in counts $g^{-1}n^{-1}cm^{-2}$ in order to determine its active mass in grams. This procedure was repeated for different temperatures, and the results can be seen in Fig.5. For the Monte Carlo method, the fit to the mass gives $m = 9.0\pm0.4$ g with a reduced $\chi^2/\nu = 1.4$. The active mass found by direct comparison with the detector calibrated by microscope is 8.2±1.4 g. The uncertainty of this method is statistical, i.e. depends on the number of neutrons generated in the Monte Carlo procedure. Although in good agreement with other loading determination techniques, an error on the neutron spectrum of the source or the possible neutron attenuation inside a detector might explain the difference between the simulation and the other methods.

Figure 6 summarizes the calibration constants obtained using these different methods. The average of the calibration constants is (0.12±0.01) counts $g^{-1}n^{-1}cm^{-2}$. Some of the methods of active mass determination presented here require the destruction of the detector. Therefore, these loading determinations are performed on some specific detectors and the active masses of detectors used for the dark matter search are calculated from their sensitivity to neutrons from the AcBe neutron source and the respective calibration constant. Since these detectors should not be subjected to a large number of events which could weaken the detector matrix, the accuracy in the loading determination is limited by statistics to about 8%.

## 5. Signal detection and efficiencies

A phase transition induces a fast expansion of the liquid droplet to a gas bubble with a factor of roughly 500 increase in volume within about 100 nsec [4]. The total energy, $W_{tot}$, which is released during this process is proportional to the droplet volume $V_{drop}$ and the stored superheat, according to the relation $W_{tot}(T) \propto V_{drop} x (P_i - P_0(T))^{3/2}$ [25]. A fraction of the released energy is converted into the acoustic energy of a pressure wave with an ultrasonic frequency spectrum, which propagates through the detector medium and is recorded by piezoelectric transducers. The transducers (Physical

Acoustics, DT15MT) are selected to match the sound emission spectrum in a frequency range from 20 kHz to 1 MHz. The transducer signals are amplified by a custom-built low-noise voltage sensitive preamplifier (variable gain up to 80 dB) and digitized by a VME based 1 MHz Flash ADC system, controlled by a PC. For the 1 litre SBD detectors two transducers are used; one sensor is glued on the middle of the exterior side of the cylindrical container wall, the other is mounted on the opposite side, 5 cm above centre and at a distance of about 10 cm from the first (Fig.1). Figure 7 illustrates the typical signal shape of a neutron-induced phase transition signal together with its Fast Fourier Transform (FFT). Usually the signals are a few milliseconds long (3 to 4 ms); their shape and FFT spectrum are dominated by the transducer response function. Despite the long signal duration, the dead time introduced by the data acquisition system itself is negligible due to the low count rate during dark matter data taking.

The observed signal amplitudes range from 40 mV to 5000 mV (with an amplifier gain of 2000) and are observed to increase by about a factor 5 with a temperature increase from $20^0$C to 45 $^0$C, the temperature operating range of SBD-1000 detectors (Fig. 8). This behaviour is predicted by the corresponding increase of the released energy $W_{tot}$ with temperature. The amplitude of the signal is proportional to the droplet size distribution; it is however also dependent on the event position in the detector due to solid angle, sound propagation and transducer response effects. Accordingly the observed distribution is a convolution of all these effects. Sound attenuation in the polymer itself is negligible.

An event is registered if at least one transducer signal is larger than the electronics threshold. Then, all channels are read out. The acceptance of the signal collection is reduced by the fraction of events missed due to amplitudes that fall below the trigger threshold. This effect is estimated for each transducer by requiring a trigger from the second transducer and by counting the number of events that fall below and above threshold for the first transducer. Since the acceptance depends on the signal amplitude it is therefore a function of the operating temperature, pressure and event location (Fig. 9).

The offline signal processing proceeds along the following steps. First, a pre-filter is applied to the data: triggers which occurred during an interval of 15 ms after a primary event are rejected to avoid re-triggering on the same event.

Second, the signals are processed by an offline filter analysis. Here they pass a filter that only retains the waveform information in the range between 40 to 130 kHz. Then, events are processed using an algorithm that yields a two-dimensional signal representation in the time-frequency domain. A threshold is applied above the level of pure electronic noise in the acoustic chain in the two-dimensional representation. The resulting signal is integrated over frequency and time to obtain a scalar value, which is assigned to the event as a quality factor, Q. Cuts on this variable allow a significant noise reduction while maintaining high overall detection efficiency.

The efficiency for this selection algorithm has been evaluated in measurements with high statistics with detectors doped with an intrinsic alpha particle activity (see section 7) and also with neutrons from an AcBe-source (data rate $\approx$ 1Hz). The filter selection depends on the amplitude distribution; it is nearly fully efficient for medium and high amplitude signals and about 50% efficient close to threshold. After weighting with the measured amplitude distribution, the trigger acceptance and the filtering efficiency are obtained as a function of temperature for each detector individually (Figs. 9 and 10). The overall detection efficiency of a detector is the product of the trigger acceptance and the Q-filter acceptance (Fig.11).

The detector operation at the pressure of operation in the SNO laboratory, where the dark matter search detector is installed, (20% above ground level pressure) implies also overall lower signal amplitudes at a given temperature compared to measurements at ground level pressure (-9% at $40^0$C and -25% at $20^0$C). Correspondingly, the total acceptance reaches its plateau for measurements in the mine at slightly higher temperatures than at the surface.

## 6. Neutron response and sensitivity to nuclear recoils

To understand the response of this type of detector to WIMP-induced nuclear recoils, one has to study the interaction of neutrons with the superheated droplets ($C_xF_y$) which leads to recoils of $^{19}$F and $^{12}$C nuclei which trigger the phase transition. In the case of nuclear recoils induced by neutrons of low energy ($E_n \leq 500$ keV), the interaction is mainly through elastic scattering on Fluorine and Carbon nuclei. Inelastic collisions are possible if the centre-of-mass kinetic energy of the neutron is higher than the first excitation level of the nuclei (1.5 and 4.3 MeV for $^{19}$F and $^{12}$C, respectively). Absorption of neutrons by the nucleus followed by ion, proton or alpha-particle emission, requires a neutron threshold energy of 2.05 MeV. Neutron absorptions may lead to the emission of γ-rays; however, droplet detectors are sensitive to γ-rays only at high temperatures of operation, outside the range of interest for dark matter search.

Assuming neutron elastic scattering on nuclei, the recoil energy, $E^i_R$, of the nucleus i is given by

$$E^i_R = \frac{2 m_n m_{N_i} E_n (1 - \cos\theta)}{(m_n + m_{N_i})^2} \tag{6}$$

where $E_n$ and $\theta$ are the incident neutron energy and the neutron scattering angle in the center-of-mass system, respectively; $m_n$ and $m_{Ni}$ are the masses of the neutron and the nucleus $i$, respectively. The recoil energy of the nucleus $i$ is largest if $\theta = 180°$:

$$E^i_{R,max} = \frac{4 m_n m_{N_i} E_n}{(m_n + m_{N_i})^2} = f_i E_n \tag{7}$$

The $f_i$ factor is the maximum fraction of the energy of the incident neutron transmitted to nucleus $i$, where $f_i = 0.19$ and $0.28$ for $^{19}$F and $^{12}$C, respectively. The ranges for $^{19}$F and $^{12}$C depend on the value of $E^i_{R,max}$ (Eq. (7)) and on their specific energy losses $dE/dx$ which are calculated from TRIM, a code calculating the transport of ions in matter [26]. At a given neutron energy, $E_n$, the recoiling nuclei, $i$ ($i = {}^{19}$F, $^{12}$C), are emitted with an angular distribution, every angle being associated to a specific recoil energy ranging between 0 keV up to a maximum energy $E^i_{R,max}$. Therefore, the nuclear recoil energy distribution, $dR_i/dE^i_R$, is completely determined by the $^{19}$F and $^{12}$C recoil angular distribution. Not all recoil energy depositions are detectable, since there exists a threshold recoil energy, $E^i_{R,th}$, below which no phase transition is triggered. $E^i_{R,th}$ depends on the temperature and pressure of operation and is related to the neutron energy threshold $E^i_{th}$ by the relation $E^i_{R,th} = f_i E^i_{th}$.

The neutron threshold energies as a function of temperature and pressure were determined by exposing detectors to mono-energetic neutrons at various temperatures and pressures. These mono-energetic neutrons were produced via the $^7Li(p,n)^7Be$ reaction with mono-energetic protons from the Tandem van de Graaff facility at the Université de Montréal. As an example Figs. 12 and 13 show the detector response (count rates) to mono-energetic neutrons of 200 and 400 keV, respectively, as a function of temperature for various pressures of operation, for a detector of 10 ml volume loaded with $C_4F_{10}$ droplets.

From such curves one can extract the threshold temperature, $T_{th}$, for a given neutron energy by extrapolating the curves down to a few degrees below the lowest point measured. From this one can infer the neutron threshold energy as a function of temperature for various pressures of operation. As can be seen from Fig.14, for a practical range of temperatures of operation, $E_{R,th}$ follows an exponential temperature dependence:

$$E_{Rth} = E_b \, e^{-K(T-T_b)} \qquad (8)$$

where $K$ is a constant to be determined experimentally and $E_b$ is the threshold energy at the boiling temperature $T_b$.

The probability, $P(E^i_R, E^i_{R,th}(T))$, that a recoil nucleus $i$ at an energy near threshold will generate an explosive droplet-bubble transition is zero for $E^i_R < E^i_{R,th}$ and will increase gradually up to 1 for $E^i_R > E^i_{R,th}$. This probability was already given in Eq (3) and expressed in terms of deposited recoil energy reads now:

$$P(E^i_R, E^i_{R,th}(T)) = 1 - \exp\left[\frac{-\alpha(E^i_R - E^i_{R,th}(T))}{E^i_{R,th}(T)}\right] \qquad (9)$$

where $\alpha$ is a parameter to be determined experimentally. Therefore, the efficiency $\varepsilon^i(E_n,T)$ that a recoil nucleus of $i$-type triggers a droplet-to-bubble phase transition at temperature $T$ after being hit by a neutron of energy $E_n$ is given by comparing the integrated recoil spectrum with and without threshold:

$$\varepsilon^i(E_n,T) = \frac{\int_{E^i_{R,\min}}^{E^i_{R,\max}} \frac{dR_i}{dE^i_R} P(E^i_R, E^i_{R,th}(T)) dE^i_R}{\int_0^{E^i_{R,\max}} \frac{dR_i}{dE^i_R} dE^i_R}. \tag{10}$$

For neutrons of energy lower than 500 keV, their collisions with $^{19}F$ and $^{12}C$ are elastic and isotropic and consequently the recoil energy distribution $dR_i/dE^i_R = 1$. Equation (10) becomes:

$$\varepsilon^i(E_n,T) = \frac{1}{E^i_{R,\max}} \int_{E^i_{R,\max}}^{E^i_{R,\min}} P(E^i_R, E^i_{R,th}(T)) dE^i_R. \tag{11}$$

Using $E^i_{R,max} = f_i E_n$ and $E^i_{R,min} = f_i E^i_{th}$, we can rewrite Eq. (11) as:

$$\varepsilon^i(E_n,T) = \left(1 - \frac{E^i_{th}(T)}{E_n} \cdot \left[1 - \exp\left\{-\alpha \frac{E_n - E^i_{th}(T)}{E^i_{th}(T)}\right\}\right] \frac{E^i_{th}(T)}{\alpha E_n}\right). \tag{12}$$

A detailed description of the neutron data has to include the energy dependence of the neutron cross section on $^{19}F$ and $^{12}C$. The energy dependence of the neutron cross section on $^{12}C$ is smooth, whereas the cross section on $^{19}F$ contains many resonances in the energy region relevant for our application [27]. One can determine the parameters $\alpha$ and $E^i_{th}$ in Eq. (12), and therefore the efficiency $\varepsilon^i(E_n,T)$, from the measured count rate, $R(E_n,T)$, of liquid-to-vapour transitions for mono-energetic neutrons of energy $E_n$ at temperature $T$:

$$R(E_n,T) = \varphi(E_n) V_l \sum_i \varepsilon^i(E_n,T) N^i \sigma^i_n(E_n), \qquad (13)$$

where $\varphi(E_n)$ is the flux of mono-energetic neutrons of energy $E_n$, $V_l$ is the volume of the superheated liquid, $N^i$ is the atomic number density of species $i$ in the liquid and $\sigma^i_n(E_n)$ the neutron cross section. The fit of Eq. (13) to the data (count rate as a function of the neutron energy) is shown in Fig. 15 for different temperatures.

From this fit we infer the efficiency $\varepsilon^i(E_n,T)$ and obtain the exponential temperature dependence for $E^i_{th}(T)$, with a temperature independent value $\alpha = 1.0 \pm 0.1$. The minimum detectable recoil energy for $^{19}F$ then becomes at 1.0 bar

$$E^F_{R,min}(T) = 0.19\, E^F_{th} = 1.55 \times 10^2 (keV)\, \exp\left[-\frac{T-20°}{5.78°}\right]. \qquad (14)$$

Combining Eqs. (3) and (14), one finds the droplet phase transition probability as a function of the recoil energy deposited by a $^{19}F$ nucleus when it is struck by a neutron. The resulting sensitivity curve, see Fig. 16, shows that the SBD-1000 detectors are more than 90% efficient at 50°C for $E_R \geq 5$keV recoils and more than 80% efficient at 40°C for $E_R \geq 25$keV recoils. Knowing the $^{19}F$ recoil spectra expected from neutralino interactions and knowing the detector response as a function of temperature (Eqs. (3) and (12)), it is then possible to determine the detector efficiency for a given neutralino mass and a given operating temperature (section 9).

In the case of a flux of poly-energetic neutrons, such as a radioactive source or neutrons from a reactor, one has to integrate Eq. (13) over the neutron energy spectrum:

$$R(T) = \int_0^{E_{n,max}} D(E_n)\varepsilon^i(E_n,T)\sigma^i_n(E_n)dE_n, \qquad (15)$$

where $D(E_n)$ is the neutron spectrum. $\sigma^i_n(E_n)$ is known from tables of neutron cross sections on $^{12}C$ and $^{19}F$ and $\varepsilon^i(E_n,T)$ has been obtained above. For a given neutron energy spectrum at low temperature, only the high energy neutrons take part in the process of liquid-to-vapour transition. The threshold energy decreases with increasing temperature, so that low energy neutrons, in addition to high energy neutrons, are also detected. Thus, for a poly-energetic neutron source, $\varepsilon^i(E_n,T)$ should increase with temperature. At high enough temperature, all neutrons in the spectrum contribute to phase transitions and $\varepsilon^i(E_n,T)$ is constant with temperature since no more neutrons are left to be detected. For a mono-energetic neutron source, there is only one strong increase of $\varepsilon^i(E_n,T)$ at the operation temperature corresponding to the energy of the phase transition and $\varepsilon^i(E_n,T)$ is constant for other temperatures. The use of droplet detectors as neutron spectrometers is based on the possibility of making the detector sensitive to different ranges of neutron energies by varying the temperature of operation. As an example, Fig. 17 shows the response of two detectors (which have different active gases and hence different neutron energy thresholds) to neutrons from an AcBe source as a function of reduced superheat.

The detector response (count rate) to mono-energetic neutrons of 200 and 400 keV as a function of operating temperature for a SBD-1000 of 10 ml volume was simulated, assuming the same energy deposition and critical length requirements as used in separate studies of the detector response to alpha particles and gamma rays (section 7 and 8). The loading was set to be 0.7% and, after analysis, the response was normalized by a multiplicative factor to fit the experimental data. This factor was interpreted as the loading correction from the initially assumed 0.7% value. The energy variation in the proton beam used to create neutrons causes the neutron energy to fluctuate around the mean energy of the beam. Consequently, the neutrons generated in the simulation were given an initial energy with Gaussian fluctuations, e.g. $\sigma = 5$ keV for 200 keV neutrons. Elastic and inelastic neutron scattering was considered for all types of nuclei using the ENDF/B-VI data library [28], in all parts of the detector. As can be seen in Figs. 18 and 19, the simulated response fits the experimental data well for $E = 200$ and 400 keV. The loading values of $f = 0.62 \pm 0.04\%$ and $f = 0.68 \pm 0.02\%$ obtained at 200 and 400 keV, respectively, are consistent with each other.

Similarly a simulation was performed for poly-energetic neutrons from an AcBe source for a 1 litre SBD-1000 detector. The virtual detector had a 4% loading in volume and 10000 neutrons were incident upon it. The Monte Carlo response is compared to data in Fig. 20.

The parameters used in the Monte Carlo calculation in order to fit the neutron curves are $l_c \approx 18 R_c$, $E^i_{R,th} = 0.28 E^i_{th}$, corresponding to Carbon nuclei triggering the phase transition at threshold, and a value of $\eta = E_c/E_{th} = 4\%$.

7. Alpha particle response

Alpha particles produced in the Freon and in the surrounding polymer matrix are the dominant intrinsic background for a dark matter search experiment based on the superheated droplet technique. Since droplet detectors are insensitive to neutron energies below a threshold value, neutrons can be effectively removed by moderating them in water or paraffin shielding. On the other hand, the detectors are sensitive to any intrinsic contamination with alpha particle emitters (other backgrounds contribute predominantly at much higher temperatures). Since the heavy salt CsCl and other ingredients used in the gel fabrication at the present stage of detector production may still contain $\alpha$-emitting contaminants such as U/Th and their daughter products, even after purification, the detector response to alpha particles has to be known with high precision.

The alpha response of our detectors was studied after adding alpha-emitting sources of known activity to the polymer matrix during the fabrication of several detectors. The detectors were placed in a thermally and acoustically insulated container, which allowed the control of the temperature with a precision of 0.1°C. The alpha response (defined as counts per emitted alpha particle per gram of active material (Freon)) was measured over a range of temperatures. Different types of detectors and radioactive sources were used. The spiked detectors were fabricated at Bubble Technology Industries (BTI), in order to avoid contamination of the Montréal production facility. During fabrication the same procedure was followed as for the non-spiked detectors used in the dark matter experiment.

Figure 21 summarizes the results of studies with detectors of the SBD-100 type. A high statistics measurement was done with a 1 litre detector with a Freon mass of (8±1) g and spiked with 20Bq of $^{232}$U. The active mass had been determined in a separate calibration with the AcBe neutron source. The detector was equipped with two piezoelectric transducers and the alpha response was measured from 4 to 35°C. At each temperature, 1000 events were recorded and a compression period of at least 4 hours took place between each measurement. Assuming that at equilibrium 5 α-particles are emitted for each $^{232}$U-decay, the count rate at high temperatures can be translated into a maximum sensitivity of about 1.4 x 10$^{-3}$ counts $\alpha^{-1}g^{-1}$.

The response curve of the 1 litre $^{232}$U-spiked detector is compared in Fig. 21 to the response of a small 10 ml SBD-100 detector spiked with 1 Bq of $^{241}$Am. The loading of this detector was very low and events were counted visually. In Fig. 21 the response of the 10 ml detector has been normalised with respect to the response curve of the 1 litre detector. For both detectors, the increase in count rate observed above 40°C is due to the onset of sensitivity to ambient gamma rays.

Other series of alpha measurements were carried out with SBD-1000 detectors. Contrary to SBD-100 detectors, for which the low threshold temperatures could not be reached in our set up, the SBD-1000 detectors allow us to explore the threshold temperature region due to their higher operating temperature. For these studies, two 1 litre SBD-1000 detectors were spiked, one with 20 Bq of $^{241}$Am and another one with 20 Bq of $^{238}$U. In the fabrication process of the $^{241}$Am-spiked detector, 27.8 ml of an Americium solution (AmCl$_3$ in 0.5 molar HCl) of known activity (0.72 Bq/ml) were added to the monomer solution. For the second detector, $^{238}$U was chosen since it mimics the natural background more adequately.

These two detectors have been calibrated with our AcBe neutron source and the active mass was found to be (7.3 ± 1.1) g for the $^{241}$Am-spiked detector and (22.1 ± 3.1) g for the $^{238}$U-spiked detector. Both detectors were equipped with two piezoelectric transducers. Figure 22 refers to the alpha response of the SBD-1000 detector spiked with $^{241}$Am (data were recorded by a PCI plug-in board). The measurements started at 15°C and went up to 50°C in steps of 1°C or 2°C, then, the temperature was set to 13°C, and lowered in steps of 2°C. At the temperature of 6°C, no events were recorded. Each measurement was preceded by a compression period lasting at least 6 hours. Approximately 500 bubbles were recorded at each temperature and the variation of the counting rate as a function of the number of created bubbles was studied. Since this variation was negligible, Fig. 22 shows the average count rate per temperature. Figure 23 refers to the alpha responses of the two spiked SBD-1000 detectors measured with a VME-based Flash ADC board as DAQ system. The data taking was automated so that the compression time, set at 10 hours, was the same for all temperatures. Approximately 500 events were recorded per temperature from 5 to 45°C, scanned in ascending and descending orders. Moreover, in order to explore the low count rate region close to threshold, the spiked detectors were shielded with 60 cm of water against ambient neutrons.

The measured alpha responses of the two SBD-1000 detectors shown in Fig. 23 are normalized according to their Freon masses (dark symbols). Only the first decay of the $^{238}$U decay-chain ($^{238}$U $\rightarrow \alpha + ^{234}$Th) was considered in the evaluation of the detector response, since, from, half-life considerations, the other members of the chain do not contribute significantly to the alpha activity. The alpha particle emitted in this decay has an energy of 4.19 MeV compared to 5.49 MeV from the decay of $^{241}$Am. The maximum response was about 1 x 10$^{-3}$ counts $\alpha^{-1}$g$^{-1}$ for the detector spiked with $^{241}$Am and 4 x 10$^{-4}$ counts $\alpha^{-1}$g$^{-1}$ for the detector spiked with $^{238}$U. The error bars shown are statistical and the three data sets (Figs 22 and 23) can be well described by a common curve fitted to the data points.

The neutron shielding used for the second measurement seems in fact to reduce the count rate at very low temperatures. However since the signal amplitudes decrease with temperature, a loss of events is expected at very low temperature and some uncertainty remains with respect to the precise location of the alpha detection threshold.

The apparent difference between the alpha sensitivities in counts $\alpha^{-1}$g$^{-1}$ of the two SDB-1000 detectors presented here might be due to different droplet size distributions of the active liquid in the gel. Indeed, on the one hand, the neutron sensitivity is proportional to the loading $f$ of the detector and independent of the droplet size distribution; on the other hand, because of their finite range in matter, alpha particles are expected to be detected with a probability which is inversely proportional to the droplet size of the active material (for the same loading). Overall, from geometric considerations, the alpha detection efficiency should scale as $\varepsilon \approx \frac{3}{4} \cdot f \cdot R_\alpha / R_d$, where $f$ is the fraction of active mass, $R_\alpha$ the alpha particle range and $R_d$ is the droplet radius, as also evinced by the MC simulations presented later. In conclusion, the droplet distribution of the detector spiked with $^{238}$U is expected to be peaked at a larger diameter than that for the $^{241}$Am-spiked detector.

In order to establish a common alpha response curve, the temperature responses of the SBD-100 and SBD-1000 detectors have been represented in Fig. 24 as a function of reduced superheat (*s*). The efficiencies of all the detectors have been adjusted for amplitude and temperature to the response of the SBD-1000 detector spiked with $^{238}$U. The combined alpha response data in Fig.24 are well represented over several orders of magnitude with the same asymmetrical sigmoid.

The alpha responses of both spiked SBD-1000 detectors were simulated, using the Geant 4.5.2 Monte Carlo code [29]. The density of the superheated droplets was varied as a function of temperature and the droplets were dispersed randomly in the gel. The ionization of low energy nuclei was taken into account, using the ICRU_R49 [30] nuclear stopping power model and the SRIM2000p [26] electronic stopping power model. The minimal energy deposition needed to trigger vaporization is known from the experimental threshold curves obtained with neutrons, assuming a head-on collision between a neutron and a nucleus inside the droplet. To maintain the consistency between neutron and alpha measurements, the simulations strongly suggest that the full energy of the recoiling nucleus (at threshold) is needed. Furthermore, to fit the alpha data, the efficiency of energy deposition $\eta = E_c/E_{th}$, i.e. the fraction of the deposited energy which appears as heat has to be, for alpha particles, larger by a factor 1.5 than for heavier ions (Carbon and Fluorine), which corresponds to $\eta \sim 6\%$.

Monte Carlo studies of the alpha response indicate that the experimental response is too high for the vaporization to be caused only by recoiling nuclei after elastic collisions between alpha particles and nuclei in the droplets. Since the Americium solution used in the spiked detector fabrication is hydrophilic and since the Freon droplets are hydrophobic, we can assume that the Americium does not diffuse into the droplets. Furthermore, the experimentally observed responses are low enough, in both the Americium and the Uranium case, to allow us to rule out any surfactant effect. Therefore the contribution of the recoiling short-range daughter nucleus can be neglected ($^{237}$Np in the case of $^{241}$Am decays), since, to be significant, this contribution would have to come from nuclei very close to the droplet surface. Thus, the phase transition must be triggered by the ionization loss of the alpha particles in the droplets.

Under the assumption that the recoiling nucleus triggering vaporization at neutron threshold is Fluorine, the *dE/dx* required to trigger a phase transition is too high to explain the efficiencies seen in the alpha case. This is not completely understood. It suggests that the minimal energy deposited at neutron threshold must be defined by the Carbon recoil. As suggested by the $^{241}$Am and $^{238}$U data above, the alpha response strongly depends on the size of the droplets dispersed in the gel. In order to prove this assumption, simulations of a 1% loaded $^{241}$Am-spiked detector were performed for five different droplet sizes (Fig.25). The maximum alpha response is shown to be inversely proportional to the droplet radius. However, the shape of the alpha response curve remains essentially unaffected by the droplet size distribution in the temperature range under investigation, as can be seen in Fig. 26.

For the simulation of the $^{241}$Am-spiked detector a loading of 0.5% was assumed, a choice validated by the neutron calibration results. Alpha particles were generated randomly in the gel, with an energy spectrum corresponding to the $^{241}$Am decay. In order for the simulation to agree with the experimental data, one has to consider that the droplet distribution within the detector is different than the one measured in an non-spiked 1 litre detector: the droplets would be approximately 1.2 times bigger in the Americium-spiked detector, causing the maximum count rate to be less than what would be expected for the measured distribution. This ad hoc supposition is to be confirmed experimentally by measuring the droplet size distribution. When this change in the droplet distribution is performed, one obtains agreement between the data and the simulation results (Fig. 23). For the Uranium-spiked detector, the loading was determined by neutron calibration to be 1.5%. The simulation was performed using the same parameters as in the Americium case, not considering the Uranium daughters, as explained earlier. As in the Americium case, one has to consider that the droplet distribution within the detector is different than the one measured in a non-spiked 1 litre detector: the droplets should be approximately 1.6 times larger in diameter. The results can be seen in Fig. 23.

Taking into account the probability function (Eq. (3)), the critical length can be obtained as a function of temperature, and the value of $\alpha$ in the probability function can be deduced from the fit to the data (see Fig. 23). The value of the critical length obtained for both simulations is $L=18R_c$ and $\alpha=1$. In the fit, the 68.3% C.L interval from Poisson statistics was used when the simulated number of events was less than 21; for higher values, the usual $\sqrt{N}$ error was employed. The hypothesis is often made that energy deposition has to occur within a spherical cavity of critical radius in order to achieve bubble nucleation. However, the simulation results show that the energy must rather be deposited over an effective track much larger than a critical diameter. Therefore, one can assume that the vapour cavity may initially extend along the particle track before quickly acquiring a spherical shape.

It is important to realize that for identical droplet distributions the alpha detection efficiencies as a function of temperature, as given by the Monte Carlo simulations, are the same for an Americium- and Uranium-spiked detector, despite the difference in alpha-particle energy (4.19 MeV for $^{238}$U and 5.49 MeV for $^{241}$Am) (Fig. 27).

## 8. Gamma response

The response of two 10 ml SBD-1000 detectors to gamma rays was studied using a $^{22}$Na source (0.7 µCi activity). The two detectors, 5.8 cm high and 1.6 cm in diameter, were each equipped with one piezoelectric transducer and placed in a temperature controlled container. The gamma ray detection sensitivity $\varepsilon_\gamma$ (counts $\gamma^{-1}$ cm$^2$) was then measured from 49.5°C to 56°C with the first detector and from 49.5°C to 70.7°C with the second (Fig. 28). Above ~72°C, spontaneous nucleation becomes significant and the detectors become intrinsically unstable. This limit of metastability of a superheated liquid is described by the homogeneous nucleation theory [15]. As emphasized in section 2, this theory predicts that the vaporization of organic liquids is taking place at an absolute temperature, in degrees Kelvin, which is about 90% of the critical temperature ($T_c$) at atmospheric pressure. According to this model one expects with decreasing temperature a very rapid exponential decrease of the spontaneous nucleation flux (s$^{-1}$cm$^{-3}$) [5]. This strong temperature dependence was confirmed in a previous experiment [5] performed close to $T \sim 0.9\ T_c$: a decrease in count rate by several orders of magnitude was observed over a temperature interval of $\Delta T = 1^o$C only. From these results one can also conclude that the process of spontaneous nucleation should not play a significant role, if a SBD-1000 detector is operated in the much lower temperature range required for detection of nuclear recoils.

The two detectors were calibrated with the AcBe n-source to determine their active mass. A loading of the order of 0.03% was found, compared to the usual loading of 0.7% for the detectors described in sections 4, 6 and 7. The response of the first detector under investigation has been normalized to the active mass of the second in Fig. 28 in order to compare their respective gamma ray sensitivities. A plateau is observed for $T \geq$ 65°C with a maximum sensitivity of $\varepsilon_{max} \sim 3.0 \times 10^{-3}$ counts $\gamma^{-1}$ cm$^2$. The measured sensitivity can be fitted over more than 6 orders of magnitude with a sigmoid function:

$$\varepsilon_\gamma = \frac{\varepsilon_{max}}{1+\exp\left(\frac{T_0 - T}{\tau}\right)}, \qquad (16)$$

with $T_0 \sim 63.8$°C and $\tau \sim 1.1$°C.

Despite the presence of the heavy CsCl salt in the gel (section 3) the average $Z$ of the detector is low ($Z \approx 5.5$) and Compton scattering is the main interaction process between γ-rays and the detector material. In particular, Compton scattering of the 1.275 MeV γ-rays of the $^{22}$Na source produces a recoiling electron with an average energy of 500 keV, which has a range of 1.1 mm in the detector medium. Knowing the average droplet radius $R$ and the detector loading $f$, the geometric efficiency that an electron of that energy hits a droplet on its trajectory is given by $\varepsilon_{geo} \approx (3/4) f L/R$. Therefore, with $R \sim 25$μm and $f = 0.03$%, one expects $\varepsilon_{geo} = 1$% and for standard detectors ($f = 0.7$%) one finds $\varepsilon_{geo} \approx 23$%. From the measured count rate, the known gamma flux, the detector loading and the Compton interaction coefficients, we can infer that on average the recoiling electrons are detected with an efficiency of 5% at the plateau, which corresponds very roughly to the estimated geometrical probability. Thus, in the plateau region Compton electrons can trigger a phase transition with high efficiency once a droplet is hit. This has also been shown to be the case for minimum ionizing cosmic ray muons [5].

Because of their very small stopping power, recoiling Compton electrons, contrary to α-particles or recoil nuclei, cannot trigger directly a phase transition in the normal temperature range of operation. The observed sensitivity to γ-rays is attributed rather to δ-rays or Auger electrons scattered off of the Compton electron tracks. These low energy electrons curl up at the end of their trajectories into localized knots of ionization and create hot spots of deposited energy of the order of a few keV or less. Figure 29 shows simulation results obtained with Geant 4.5.2 for the energy spectra of δ-rays emitted on tracks of 500 keV and 5 keV electrons and of 1 GeV muons, respectively, sent through the detector material. The δ-ray spectrum is independent of the primary electron and muon energy and 50% of the emitted δ-rays deposit energies smaller than 0.05 keV.

Using this δ-ray spectrum in the simulation analysis code with the parameters described in Section 6, one finds a detector response which corresponds roughly to the measured one, as can be seen on Fig. 30. In this figure, the simulated response was normalized to the measured one at the plateau and a shift of 2°C in temperature was performed. This shift could be accounted for by a better calibration of the temperature sensor, which is underway. A more complete simulation of the gamma response is planned, but the results obtained so far indicate that the overall γ-ray sensitivity of our detector can be understood as being proportional to the product of the Compton cross section and the δ-ray emission probability.

The γ-ray counting efficiency of the SBD-1000 and SBD-100 detectors in the plateau region has been studied in more detail with different γ-ray sources. Figure 31 shows the number of bubbles triggered per incident γ, as a function of the γ- ray energy and normalized to a loading of 0.7%. As can be seen from Fig. 31, the counting efficiency for γ-rays is practically constant from ~100 keV up to 1.3 MeV. This can be understood if one considers that because of the low average-Z (Z ≈ 5.5) of the detectors, the Compton scattering cross section is dominant from ~ 400 keV up to ~ 5 MeV and changes only slightly over this energy region. Moreover since the specific energy loss of the Compton electrons is also basically constant in this region and the δ-ray emission probability on the electron track itself is proportional to the *dE/dx*, we expect an overall flat detection efficiency, with a linear rise below ~ 500 keV.

## 9. Response to neutralino induced recoils

There is strong evidence that a large fraction of the matter in the Universe is non-luminous and non-baryonic, exotic in nature [31]. On galactic scales, the evidence for dark matter comes from the rotation curves of spiral galaxies. To explain the observed curves a halo of dark matter that extends far beyond the luminous disk has to exist. The most promising candidate for dark matter from the particle physics point of view is the neutralino, an electrically neutral, weakly interacting massive particle (WIMP), which occurs naturally in supersymmetric theories with a mass in the range from 10 GeVc$^{-2}$ to 1 TeVc$^{-2}$. In particular, in our galaxy, the neutralinos are assumed to follow an isotropic Maxwellian velocity distribution of the form $f(v) = v^2 e^{-(v+v_E)^2/v_0^2}$, where $v_0$ = 230 kms$^{-1}$ is the velocity dispersion of the dark matter halo and $v_E$ = 244 kms$^{-1}$ is the velocity of the earth relative to the dark matter distribution, ignoring the effect of a $\pm 20$ kms$^{-1}$ annual variation. The local neutralino mass density at the position of the solar system is assumed to be 0.3 GeVcm$^{-3}$ and the velocity distribution has a maximum at the escape velocity of matter of the Milky Way, $v_{esc}$ = 600 kms$^{-1}$. Most of these parameters are model dependent and known with large uncertainties only. However, following a suggestion presented in [32] these values are generally adopted as bench marks to allow the comparison of results and sensitivities of different experiments.

The elastic cross section of neutralino scattering off nuclei in the detector is of electroweak strength and, in the zero-momentum transfer limit, has the general form given by

$$\sigma_A = 4 G_F^2 \left( \frac{M_\chi M_A}{M_\chi + M_A} \right)^2 C_A \qquad (17)$$

where $G_F$ is the Fermi constant, $M_\chi$ and $M_A$ are the mass of the neutralino and the detector nucleus, respectively [11]. $C_A$ is an enhancement factor which depends on the form of the neutralino interaction: i) coherent or spin-independent (SI) interactions are described by an effective scalar coupling between the neutralino and the nucleus with $C_A$ given by $C_A^{SI} = \frac{1}{4\pi} \left[ Z f_p + (A-Z) f_n \right]^2$ where $f_p$ ($f_n$) are the neutralino couplings to the proton (neutron). For equal coupling to neutrons and protons this cross section is proportional to $A^2$; ii) incoherent or spin-dependent (SD) interactions involve axial currents and couple the spin of a neutralino to the total spin of the nucleus. Since the spin of the nucleus can be carried by protons and neutrons, the enhancement factor is of the form $C_A^{SD} = \frac{8}{\pi}[a_p <S_p> + a_n <S_n>]^2 \frac{J+1}{J}$, where $<S_p>$ ($<S_n>$) are the expectation values of the proton (neutron) spin in the target nucleus, $a_p$ ($a_n$) are the effective proton (neutron) coupling strengths and $J$ is the total nuclear spin. Due to their $^{19}$F content, PICASSO SBD detectors are especially suitable to search for spin-dependent neutralino interactions, since $^{19}$F is a spin-$1/2^+$ isotope and has a very favorable spin dependent cross section [33].

Again in order to compare different experiments using different target nuclei the convention is followed to convert results and sensitivities involving WIMP cross sections on target nuclei, into WIMP-proton ($\sigma_p$) and WIMP neutron ($\sigma_n$) cross sections, respectively. To do this, one follows the procedure in [34] and assumes that all events are either due to WIMP-proton or WIMP-neutron elastic scatterings in the nucleus, i.e $a_n = 0$ or $a_p = 0$, respectively, and one has $\sigma_{p,n} = \sigma_A (\mu_{p,n}^2/\mu_A^2) C_{p,n}/C_{p,n(A)}$ where $\mu_{p,n}$ is the WIMP-nucleon reduced mass and $C_{p(F)}$ and $C_{n(F)}$ are the proton and neutron contributions to the total enhancement factor of $^{19}$F. The latter in turn are related to the $a_p$ and $a_n$ couplings by $C_{p,n(A)} = (8/\pi)a_{p,n}^2 <S_{p,n}>^2 (J+1)/J$. $C_p$ and $C_n$ are the enhancement factors for scattering on individual protons and neutrons. For $^{19}$F, the values for these ratios are $C_{p(F)}/C_p = 0.778$ and $C_{n(F)}/C_n = 0.0475$ and for $^{23}$Na $C_{p(Na)}/C_p = 0.137$ and $C_{n(Na)}/C_n = 8.89 \times 10^{-4}$, respectively [34,35].

To detect neutralinos, one measures the energy of nuclear recoils arising from neutralino elastic scattering off nuclei in the detector. With the above values for neutralino masses and velocities, recoil energies are expected to be less than 100 keV for all detector materials. In the PICASSO case, we are interested in spin-dependent (SD) neutralino interactions on $^{19}F$ only. Being light nuclei, $^{19}F$ and $^{12}C$ have negligible spin-independent interactions, see for instance [36,37].

In order to calculate the expected $^{19}F$ recoil spectrum for interactions with neutralinos in our galactic halo we use the formalism described in [32]. The observable rate as a function of neutralino mass, cross section and operating temperature is then given by

$$R_{obs}(M_\chi, \sigma_{SD}, T) = \int_0^\infty P(E_R, E_{RTh}(T)) \frac{dR(v_E, v_{esc}=\infty)}{dE_R} dE_R \qquad (18)$$

where $P(E_R, E_{RTh}(T))$ is the recoil energy threshold function of our SBD detectors as a function of temperature determined by neutron calibrations and given by Eq. (9). $dR/dE_R$ is the neutralino induced recoil energy spectrum of $^{19}F$ nuclei assuming $v_E = 244\,km\,s^{-1}$ and with $v_{esc} = \infty$, since the truncation effect on the count rate assuming $v_{esc} = 600$ kms$^{-1}$ is only at the $10^{-3}$ level. For practical purposes the recoil spectrum can then be well approximated by an exponentially falling distribution of the form [32]:

$$\frac{dR}{dE_R} \approx c_1 \frac{R_0}{\langle E_R \rangle} F^2(E_R) \exp\left(-\frac{c_2 E_R}{\langle E_R \rangle}\right) \qquad (19)$$

where $\langle E_R \rangle = 2 M_A M_\chi^2/(M_A+M_\chi)^2 \langle v_\chi^2 \rangle$ is the mean recoil energy; $F^2(E_R)$ is a nuclear form factor close to 1 for a light nucleus like Fluorine and for small momentum transfer; the constants $c_{1,2}$ describe the effect of the earth's velocity relative to the halo ($c_{1,2}=1$ for $v_E = 0$; $c_1 = 0.75$ and $c_2 = 0.56$ for $v_E = 244$ kms$^{-1}$); $R_0$ is the total rate in counts kg$^{-1}$day$^{-1}$ assuming zero momentum transfer ($\sigma$=const.) and $v_E = 0$ given by

$$R_0(counts\ kg^{-1}d^{-1}) = \frac{403}{A_T \cdot M_\chi}\left(\frac{\sigma_{SD}}{pb}\right)\left(\frac{\rho_\chi}{0.3 GeVcm^{-3}}\right)\left(\frac{<v_\chi>}{230 kms^{-1}}\right) \quad (20)$$

where $A_T$ denotes the atomic mass of the target atoms, $\rho_\chi$ is the mass density of neutralinos, $\sigma_{SD}$ the neutralino cross section under investigation and $<v_\chi>$ the relative average neutralino velocity.

Combining the $^{19}$F recoil spectra expected from neutralino interactions (Eq. (14)) and the measured detector threshold for Fluorine recoils at a given operating temperature as a function of temperature (Fig. 16), we can determine the neutralino detection efficiency as a function of neutralino mass and operating temperature [17,22]:

$$\varepsilon(M_\chi, T) = \frac{\int_0^\infty P(E_R, E_{RTh}(T))\frac{dR}{dE_R}dE_R}{\int_0^\infty \frac{dR}{dE_R}dE_R} = \frac{c_2}{c_1}\frac{1}{R_0}\int_0^\infty P(E_R, E_{RTh}(T))\frac{dR}{dE_R}E_R$$

(21)

The resulting detection efficiencies for SBD-1000 detectors operated at mine pressure are shown in Fig. 32 for neutralinos of masses in the range from 10 to 500 GeVc$^{-2}$. Finally the observable neutralino count rate as a function of temperature, neutralino mass and cross section is given by

$$R_{obs}(M_\chi, \sigma_{SD}, T) = \frac{c_1}{c_2}R_0(M_\chi, \sigma_{SD}) \cdot \varepsilon(M_\chi, T) = 1.34 \cdot R_0(M_\chi, \sigma_{SD}) \cdot \varepsilon(M_\chi, T) \quad (22)$$

where the numerical factor of 1.34 describes the effect of $v_E$ and all detector specific properties which enter via $\varepsilon(M_\chi, T)$. The expected count rate for SBD-1000 detectors at mine pressure is shown in Fig. 33, considering a neutralino-Fluorine cross section of 100 pb and neutralino masses of 10, 25, 50, 100 and 500 GeVc$^{-2}$, respectively.

A first set of low background SBD-1000 detectors of the kind described here has been installed at the Sudbury Neutrino Observatory to search for spin-dependent WIMP induced interactions on $^{19}$F. The results from three superheated droplet detectors with a total active mass of 19.4 ± 1.0 g of $^{19}$F and an exposure of 1.98 ± 0.19 kgd were analysed and reported in [1]. A combined fit of alpha background and neutralino response to the data allowed the ruling out of any positive evidence for WIMP induced nuclear recoils, yielding a 90% C.L. upper limit of 1.31 pb on protons and 21.5 pb on neutrons for a WIMP mass $M_{WIMP}$ = 29 GeVc$^{-2}$. The sensitivity of the experiment is presently limited by alpha emitting contaminants and its small active mass. The next step of the experiment is in preparation, with 2 kg of active mass, detector modules of 4.5 litre volume, hydraulic recompression, event localization capability and improved purification and fabrication techniques.

## 10. Conclusions

For the first time we present results of a comprehensive systematic study of the response of superheated liquid droplet detectors used in the PICASSO dark matter experiment to different kinds of radiation. All our results could be described in a consistent manner with energy depositions ranging from x-rays in the keV range to nuclear recoils in the several MeV range. Since the energies of nuclear recoils following dark matter neutralino interactions can be easily reproduced in neutron scattering, we performed extensive studies with neutron beams and sources. Of particular interest for dark matter applications is the dependence of the detector threshold for $^{19}$F recoils on the SBD operating temperature. Combining these threshold measurements with the $^{19}$F recoil spectra expected from neutralino interactions, one can determine the neutralino detection efficiency and finally the expected count rate for a given interaction cross section and neutralino mass at different temperatures.

The heavy salt and other ingredients, mixed in the gel at the present stage of detector fabrication, still introduce contaminants which are α-emitters, such as U/Th and their daughter nuclei. Therefore, the α-background is the main background in the dark matter search. Other potential backgrounds, such as γ-rays and cosmic ray muons (or other mips), contribute to the detector signal predominantly at higher temperatures. The alpha response of SBD detectors was studied by adding alpha-emitting sources of known activity to the polymer matrix during the fabrication of several detectors. A consistent description of the alpha response for a series of alpha spiked detectors has been obtained with different active detector liquids and therefore different operating temperatures, if represented in terms of reduced superheat (with a critical temperature at 90% of its tabulated value). The range of alpha sensitivity coincides with the range of temperatures where droplet detectors are sensitive for neutralino induced recoils, but the shape of the response curves are different. Precise knowledge of the alpha response is therefore important in order to increase the sensitivity for neutralino detection in the presence of alpha emitting contaminants. The measured alpha response of the detectors is well reproduced in simulations using a consistent set of variables that parameterize the underlying model of recoil energy threshold and energy deposition (theory of Seitz).

Finally, the response of SBD detectors to γ-rays has been studied with sources and simulations. The observed high sensitivity for Compton electrons and mips could be attributed to δ-rays scattered off the primary particle tracks, which curl up at the end of their trajectories, creating hot spots of deposited energy of the order of keV. Simulations based on the same theoretical parameters reproduce the neutron data using the same energy deposition and critical length requirements as found in the case of alpha particle and γ-rays induced events. Thus from these studies emerges a complete characterization of the response of detector modules already in use for a dark matter search by PICASSO at SNO.

## Acknowledgements

This work has been funded by the Canadian National Science and Research Council (NSERC) and partial support has been obtained from the Laboratory of Advanced Detector Development (LADD) in the framework of the Canada Foundation for Innovation (CFI). The authors from Indiana University South Bend wish to thank the IUSB R&D committee, and the office of the Dean of the College of Liberal Arts and Sciences for partially funding this work. The skilful technical support of J. Berichon and G. Richard (UdeM) during all phases of the project was very much appreciated. Many thanks go to G. Azuelos (UdeM) for helpful discussions. We thank J. Bocan, S. Pospisil and I. Stekl (Prague) for measuring the Rn diffusion in PICASSO container walls. The support and hospitality of the Sudbury Neutrino Observatory and its staff are gratefully acknowledged.

Fig. 1: A 1-litre volume droplet detector module equipped with piezoelectric transducers glued on the surface. The container is capable of holding pressures up to 10 bars. Typical $C_xF_y$ gas loading presently achieved for these detectors is in the 5-10 g/litre range.

Fig. 2: Evolution of the energy threshold for $^{19}$F recoils as a function of temperature (1.23 bars). At threshold the $^{19}$F recoil detection efficiency rises gradually and the broken (continuous) lines indicate 50% (80%) detection efficiencies, respectively. From kinematical considerations, WIMP induced recoil energies are smaller than 100 keV and become detectable above 30°C; at around 15° C the detector becomes sensitive to alpha particles from U/Th contaminations. Above 55°C recoils with energies below one keV can be detected, but at the same time the detector becomes sensitive to γ-rays, minimum ionizing particles due to associated δ- and Auger electrons; at the foam limit the detector becomes intrinsically unstable.

Fig. 3: The background count rate ($g^{-1}d^{-1}$) as a function of the detector fabrication date, from the time when no purification was performed before fabrication until the time when all ingredients were purified.

Fig. 4: Comparison between the droplet diameter distributions (in μm) measured for a 1 litre detector (dotted) and a 10 millilitre (plain curve) detector.

Fig. 5: Active mass determination for a 1 litre detector using a Monte Carlo simulation. Data is simulated at different temperatures with a neutron energy spectrum corresponding

to that of an AcBe source. A loading of 4% has been assumed. The shaded region corresponds to the experimentally determined mass.

Fig. 6: Calibration constants (C) of several detectors obtained with four techniques of measurement: neutron beam (♦), Monte Carlo simulation (□), visual determination (▲) and direct measurement (○). The straight line indicates the average count rate (0.12±0.01) $g^{-1}n^{-1}cm^2$ of the calibration constants.

Fig. 7: Phase transition signal recorded by a piezo-electric transducer [amplitude (mV) as function of time (μs)]. The insert shows the corresponding Fast Fourier Transform [rms amplitude (mV) as function of frequency (kHz)].

Fig. 8: Observed signal amplitude dependence on temperature: (+: 20°C, 25°C average), (*:35°C, 40°C). Overflow counts at high temperature have been suppressed.

Fig. 9: Trigger acceptance for data taken with two SBD-1000 detectors spiked with alpha emitters with no source ($^{241}$Am (♦) and $^{238}$U (Δ)) and in front of a neutron source ($^{241}$Am (◊) and $^{238}$U (▲)). The acceptance is constant (~95%) over the temperature range where a dark matter signal is expected.

Fig. 10: Efficiency of the Q-filter for data taken with two SBD-1000 detectors spiked with alpha emitters with no source ($^{241}$Am (♦) and $^{238}$U (Δ)) and in front of a neutron

source ($^{241}$Am (◊) and $^{238}$U (▲)). The efficiency is constant (~90%) over the temperature range where a dark matter signal is expected.

Fig. 11: Overall detection efficiency of the Q-filter for data taken with two SBD-1000 detectors spiked with alpha emitters ($^{241}$Am (♦) and $^{238}$U (Δ)). A sigmoid fit to the data is also shown to guide the eye. The efficiency is constant (85%) over the temperature range where a dark matter signal is expected.

Fig. 12: Detector response to 200 keV neutrons as a function of temperature ($^o$C) at various pressures (from the left to the right: 1.01, 1.36 and 1.70 bars, respectively). The 10 ml detector is loaded with 100% $C_4F_{10}$ gas. The 200 keV neutrons used for these measurements were obtained from $^7$Li(p,n)$^7$Be reactions at the tandem facility of the Université de Montréal [27].

Fig. 13: Detector response to 400 keV neutrons as a function of temperature ($^o$C) at various pressures (from the left to the right: 1.01, 1.36, 1.70 and 2.05 bars, respectively). The 10 ml detector is loaded with 100% $C_4F_{10}$ gas. The 400 keV neutrons used for these measurements were obtained from $^7$Li(p,n)$^7$Be reactions at the tandem facility of the Université de Montréal [27].

Fig. 14: Neutron threshold energy ($E_{Rth}$) in keV as a function of temperature for various operating pressures (from the bottom to the top: 1.01, 1.36 and 1.70 bars, respectively) [27].

Fig. 15: Response of a SBD-100 droplet detector to a monoenergetic beam of neutrons for different temperatures: T=10°C (dashed line), T=15°C (dotted line) and T=20°C (plain line). Using the known tabulated neutron cross sections on $^{12}$C and $^{19}$F, the fit to the data for different temperatures gives an exponential temperature dependence for $E^i_{th}(T)$, and the efficiency $\varepsilon^i(E_n,T)$ is obtained with $\alpha = 1.0 \pm 0.1$. In the ordinate, the count rate is in arbitrary units.

Fig. 16: Probability for phase transition in a SBD-1000 detector as a function of deposited recoil energy ($E_R$) and temperature for $^{19}$F nuclei.

Fig. 17: The response of two detectors, a SBD-100 (♦) and a SBD-1000 (▲), to neutrons from an AcBe source as a function of reduced superheat, s. The use of the reduced superheat parameter allows the unification of the response of the two detectors. With parameter s, the response for different gases is computed with $T_{ceff} = 0.9\ T_c$ [27].

Fig. 18: Simulation of the SBD-1000 response (count rate), at a pressure of 1.01 bars, to 200 keV neutrons as a function of temperature (°C) compared with the experimental data. The volume of the detector is 8 ml. The simulated response gives a loading of (0.62±0.04)%.

Fig. 19: Simulation of the SBD-1000 response (count rate), at a pressure of 1.01 bars, to 400 keV neutrons as a function of temperature (°C) compared with the experimental data.

The volume of the detector is 8 ml. The simulated response gives a loading of (0.68±0.02)%.

Fig. 20: Neutron detection efficiency ($g^{-1}n^{-1}cm^2$) as a function of temperature for an AcBe source. The shaded region corresponds to the simulation results, (*) represent the experimental data.

Fig. 21: The alpha response ($\alpha^{-1}g^{-1}$) of two SBD-100 detectors spiked with 1 Bq of $^{241}$Am (♦) and 20 Bq of $^{232}$U (Δ) as a function of temperature (°C). The increase in the counting rates observed above 40°C is due to gamma rays.

Fig. 22: The alpha response ($\alpha^{-1}g^{-1}$) as a function of temperature (°C) of a SBD-1000 detector spiked with 20Bq of $^{241}$Am normalized by its active mass. Data are fitted with an asymmetrical sigmoid.

Fig. 23: The response ($\alpha^{-1}g^{-1}$) as a function of temperature (°C) of two SBD-1000 detectors normalized by their active mass: $^{241}$Am-spiked (♦) and $^{238}$U-spiked (▲). For the same amount of active material the $^{241}$Am-spiked detector is twice as sensitive to alphas as the $^{238}$U-spiked detector. The data are fitted with the previous asymmetrical sigmoid adjusted in amplitude and slightly shifted in temperature. The results of the simulations are shown for both $^{241}$Am-spiked (◊) and $^{238}$U-spiked (Δ) detectors.

Fig. 24: The response ($\alpha^{-1}g^{-1}$) of SBD-100 (spiked with $^{241}$Am (○) and spiked with $^{232}$U (Δ)) and SBD-1000 detectors (spiked with $^{241}$Am series 1 (+) and series 2 (×) and spiked with $^{238}$U (●)) as a function of the reduced superheat s. The data are fitted over several orders of magnitude with the asymmetrical sigmoid.

Fig. 25: Simulated alpha detection response ($\alpha^{-1}$) of a 1% loaded SBD-1000 detector spiked with 20 Bq of $^{241}$Am as a function of droplet radius for three different temperatures: 20°C (▪), 30°C (•) and 40°C (*). The alpha detection response is inversely proportional to the droplet size.

Fig. 26: Ratio, as a function of temperature (°C), between the response obtained from simulations from three detectors, each containing droplets of a different radius (10 μm (x), 15 μm (•) or 30 μm (*)), and the response of a simulated detector containing a droplet size distribution as measured in one 1 litre detector. The responses of all detectors had been previously normalized to 1 at maximal detection response. As the ratio is compatible with 1 from 20°C to 50°C, one can conclude that the droplet size distribution does not influence the alpha curve shape within this range.

Fig. 27: Comparison between the simulated alpha detection response as a function of temperature (°C), for the americium (plain line) and uranium spiked (dashed line) detectors, considering the same droplet size distribution in both cases.

Fig. 28: The measured gamma-rays detection sensitivity, $\varepsilon$ ( $\gamma^{-1}$ cm$^2$), as a function of temperature ($^o$C) for two 10 ml SBD-1000 detectors(first detector (▪) and second detector (*)) using a $^{22}$Na source. The response can be approximated by a sigmoid function using Eq. (16) (dashed line).

Fig. 29: Energy spectrum of δ-rays emitted on tracks of 1 GeV muons (o), 500 keV electrons (+), and 5 keV electrons (*) sent through the detector material. The line represents a fit to the combined spectra.

Fig. 30: The simulated detector response to gammas as a function of temperature (✣) obtained with the δ-ray spectrum of Fig. 29 after a shift of 2°C. The dotted line corresponds to the measured response to which the simulated response was normalized in the plateau region.

Fig. 31: The counting efficiency for γ-rays, $\varepsilon(\gamma^{-1})$, is shown as a function of the γ-ray energy (MeV) and normalized to a loading of 0.7% : (♦) are results from [5], (□) are the results from the measurement of two SBD-1000 detectors exposed to a 0.7 μCi $^{22}$Na γ-source (see Fig. 28).

Fig. 32: Detection efficiency of a SBD-1000 detector at mine pressure (1.2 bars) for the detection of neutralino of masses 500, 100, 50, 25 and 10 GeVc$^{-2}$.

Fig. 33: Dark matter count rate ($g^{-1}d^{-1}$) of SBD-1000 detectors at mine pressure (1.2 bars) considering a neutralino-fluorine cross section of 100 pb and neutralino masses of 10, 25, 50, 100 and 500 $GeVc^{-2}$.

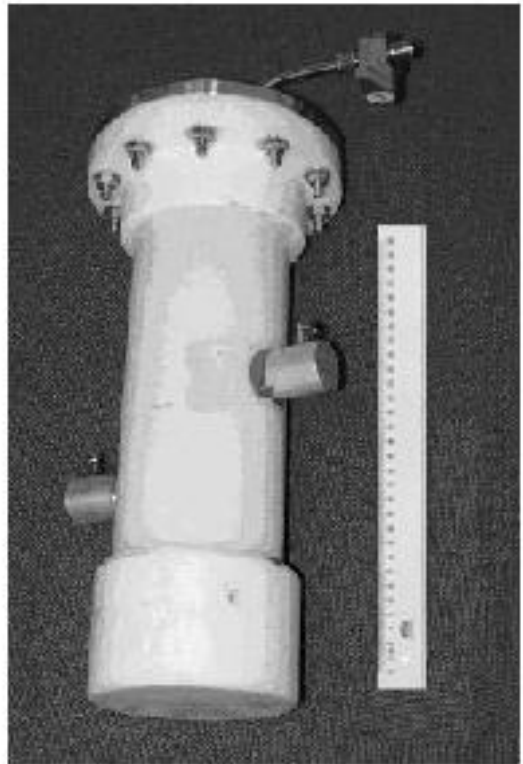

Figure 1

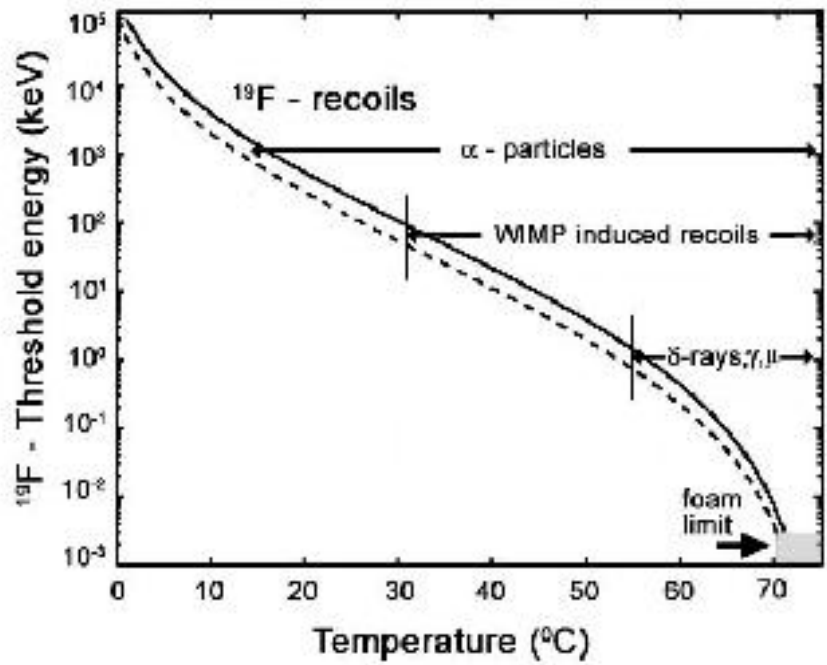

Figure 2

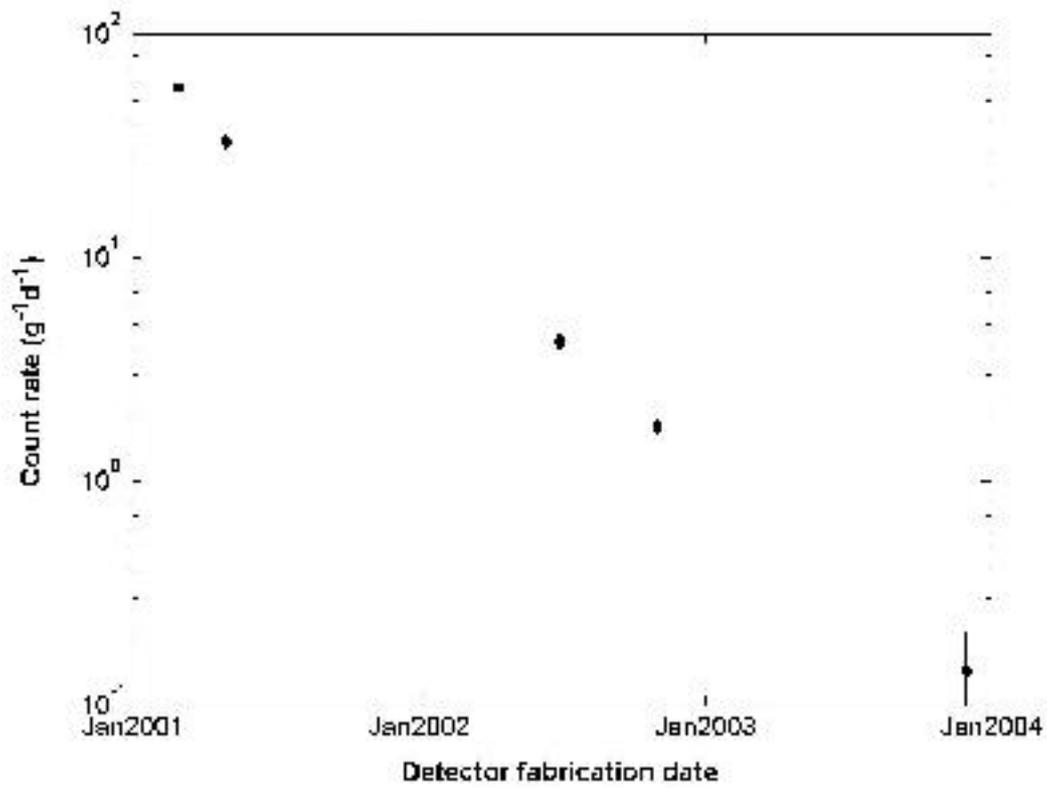

Figure 3

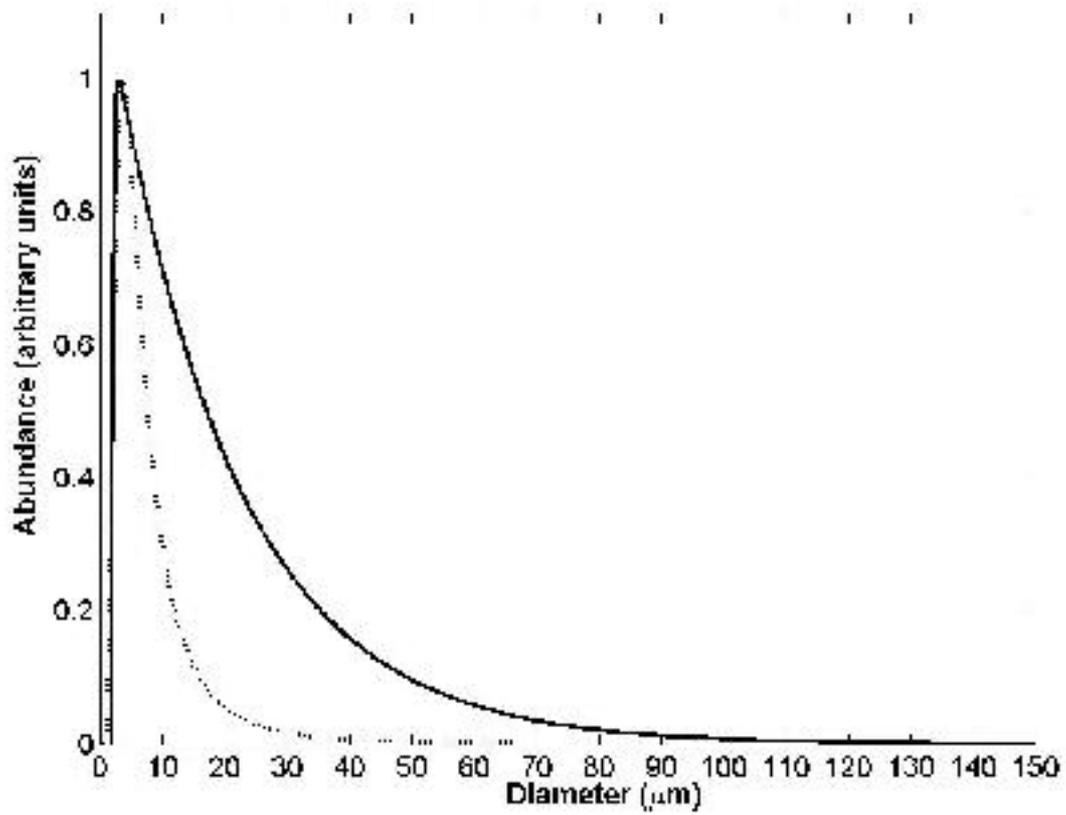

Figure 4

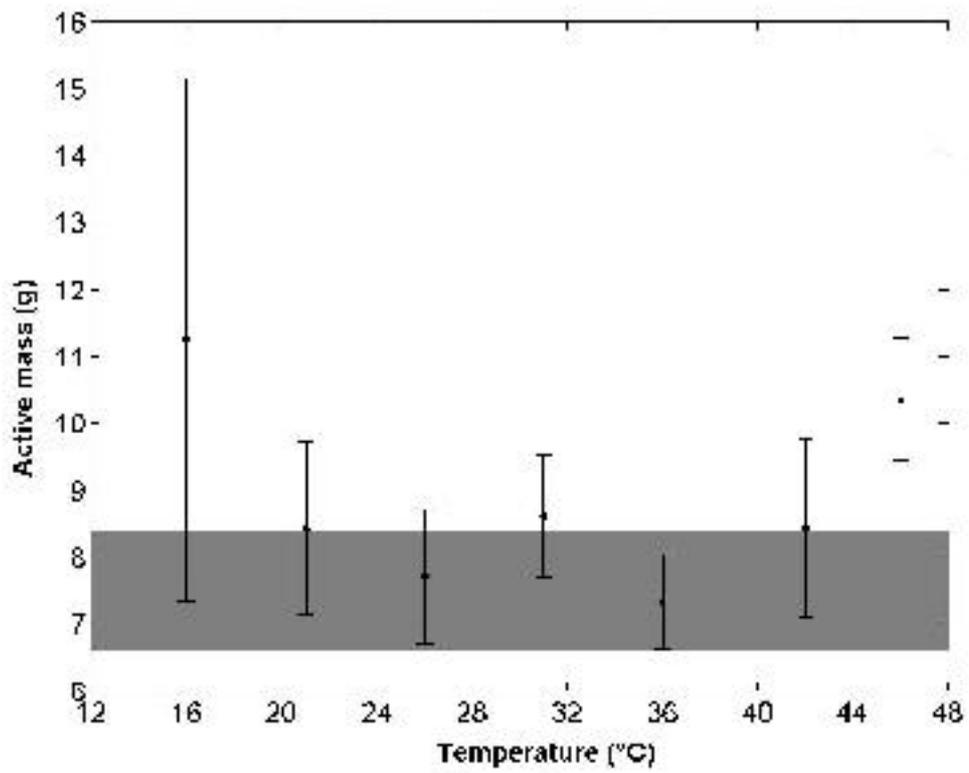

**Figure 5**

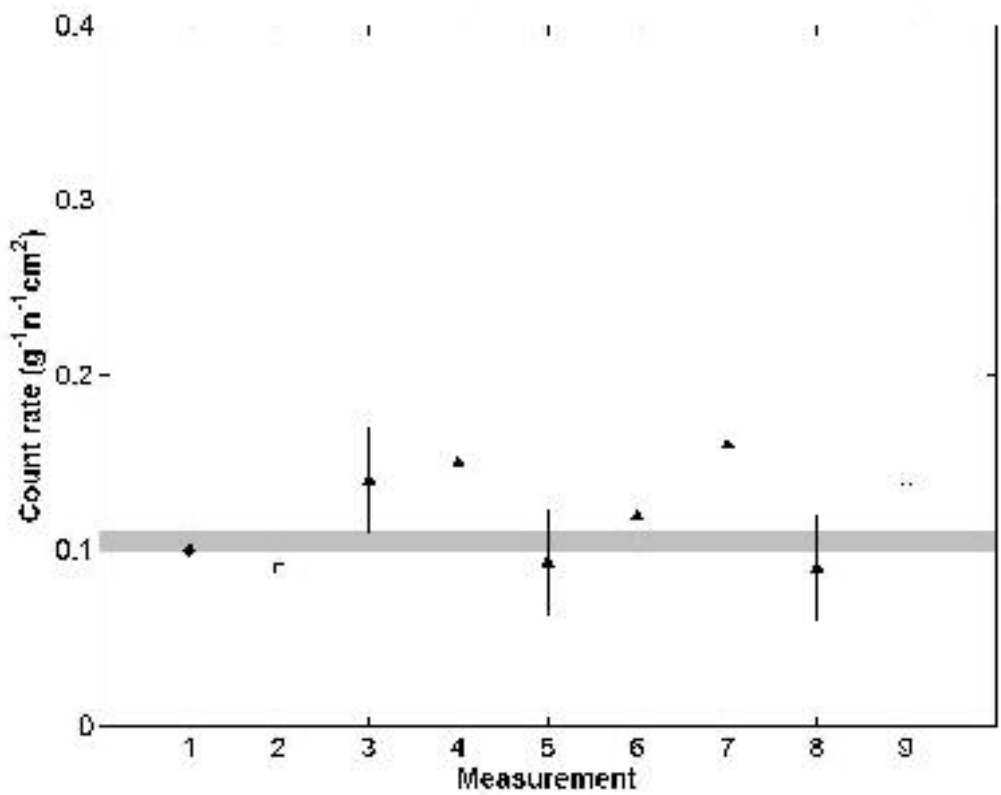

**Figure 6**

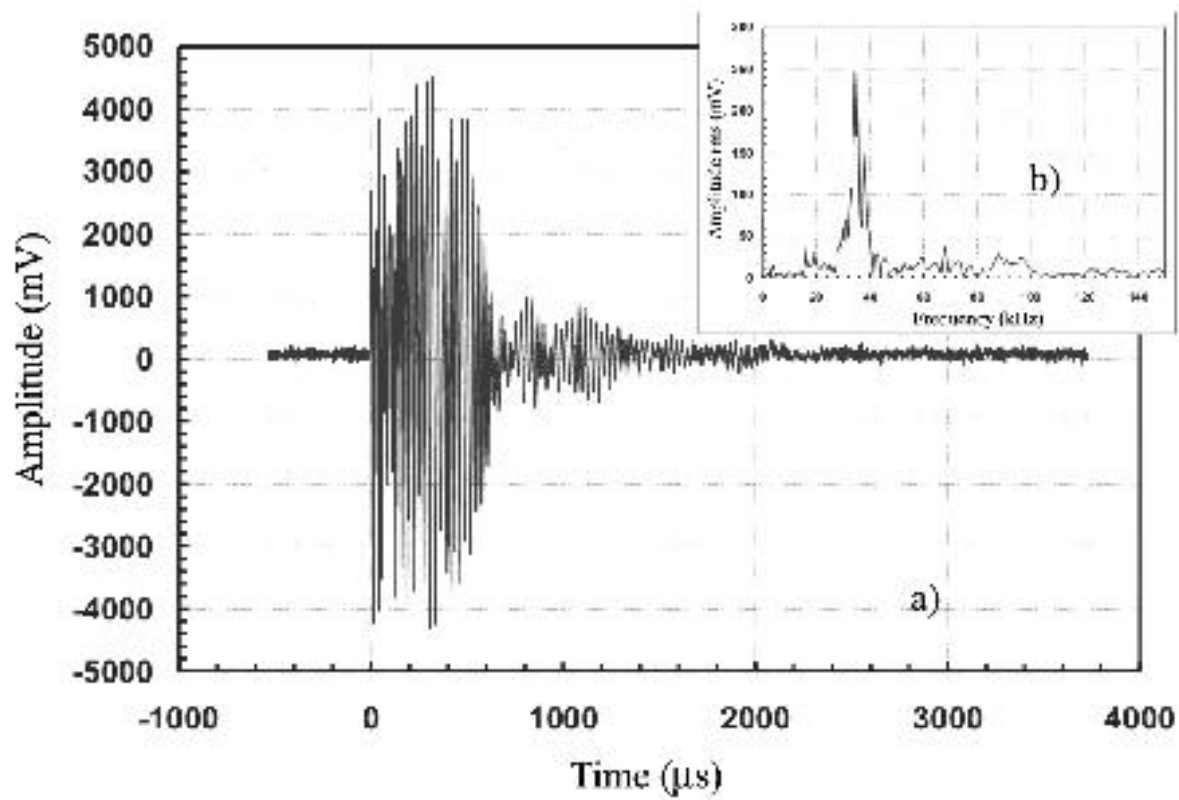

**Figure 7**

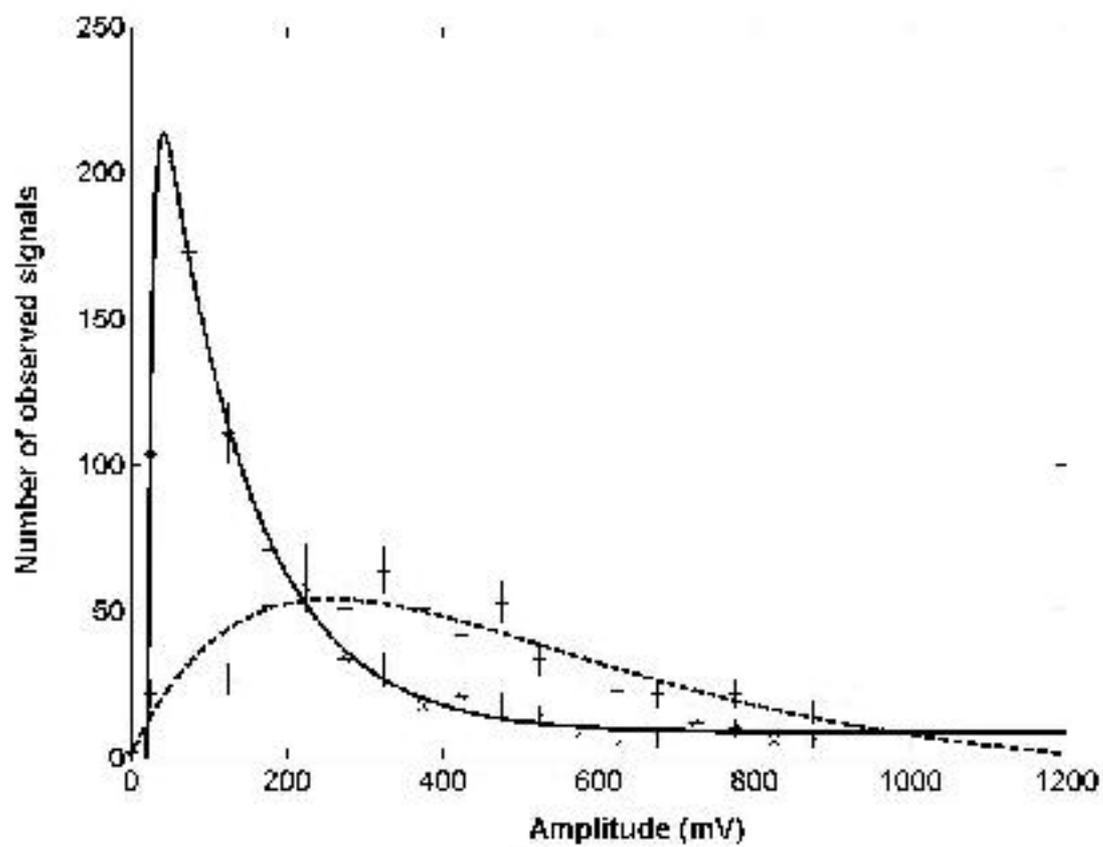

**Figure 8**

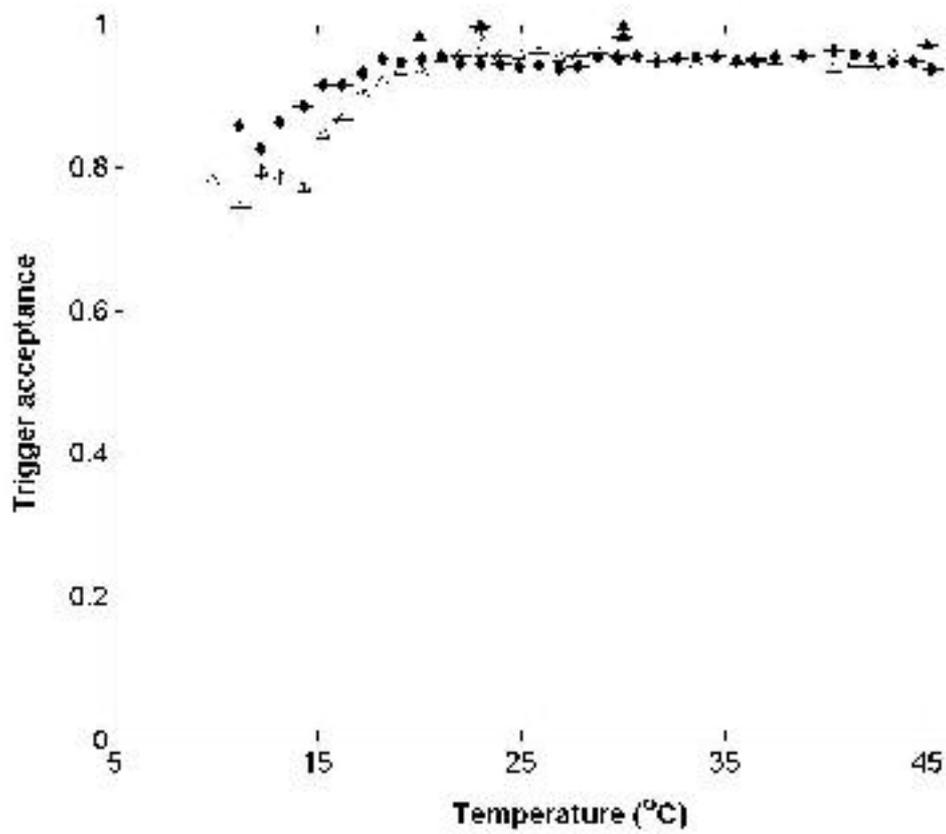

Figure 9

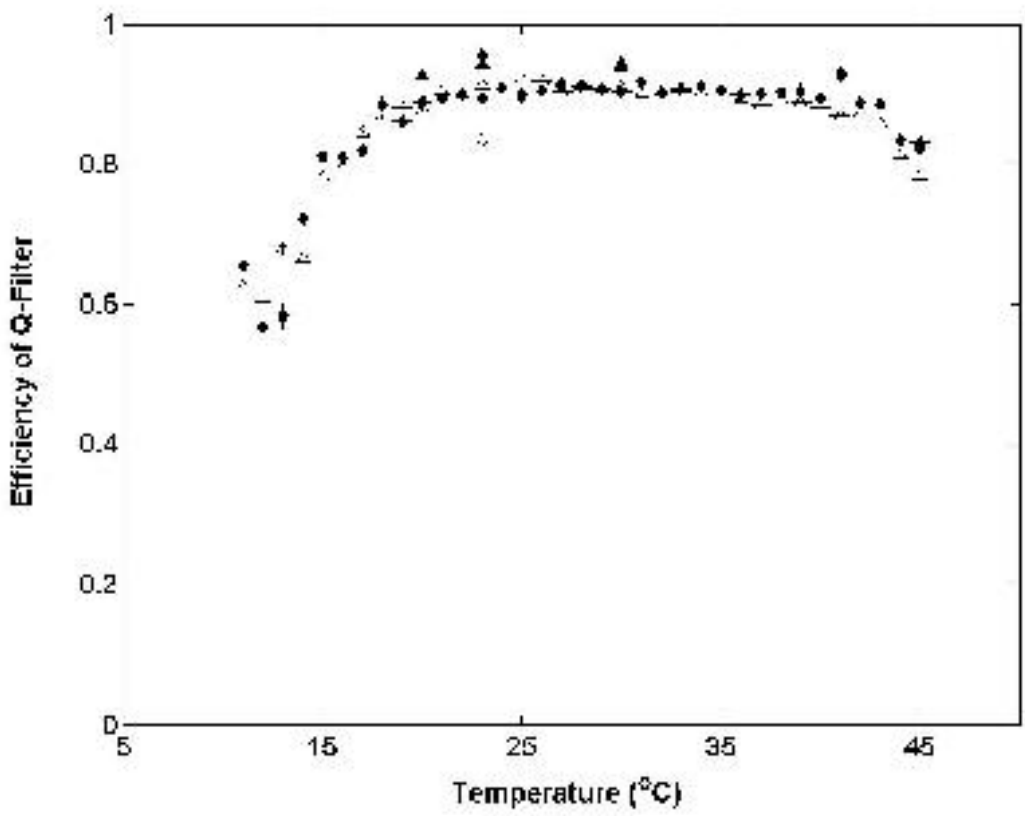

Figure 10

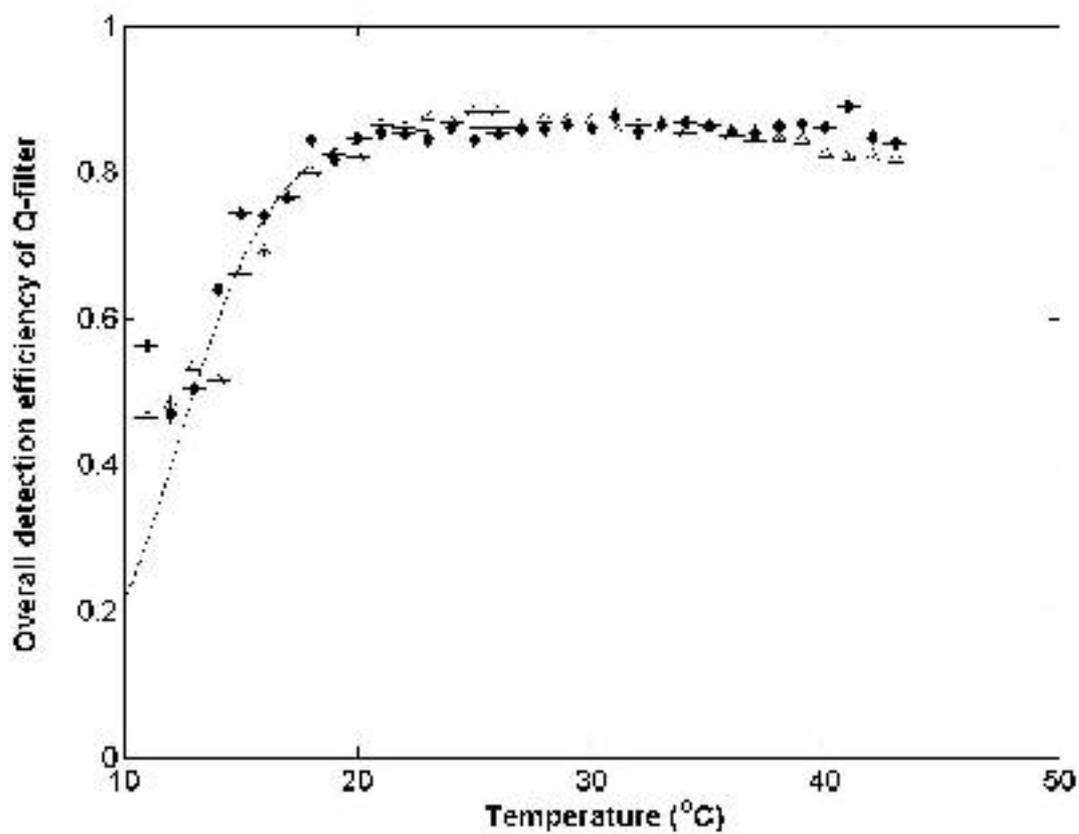

Figure 11

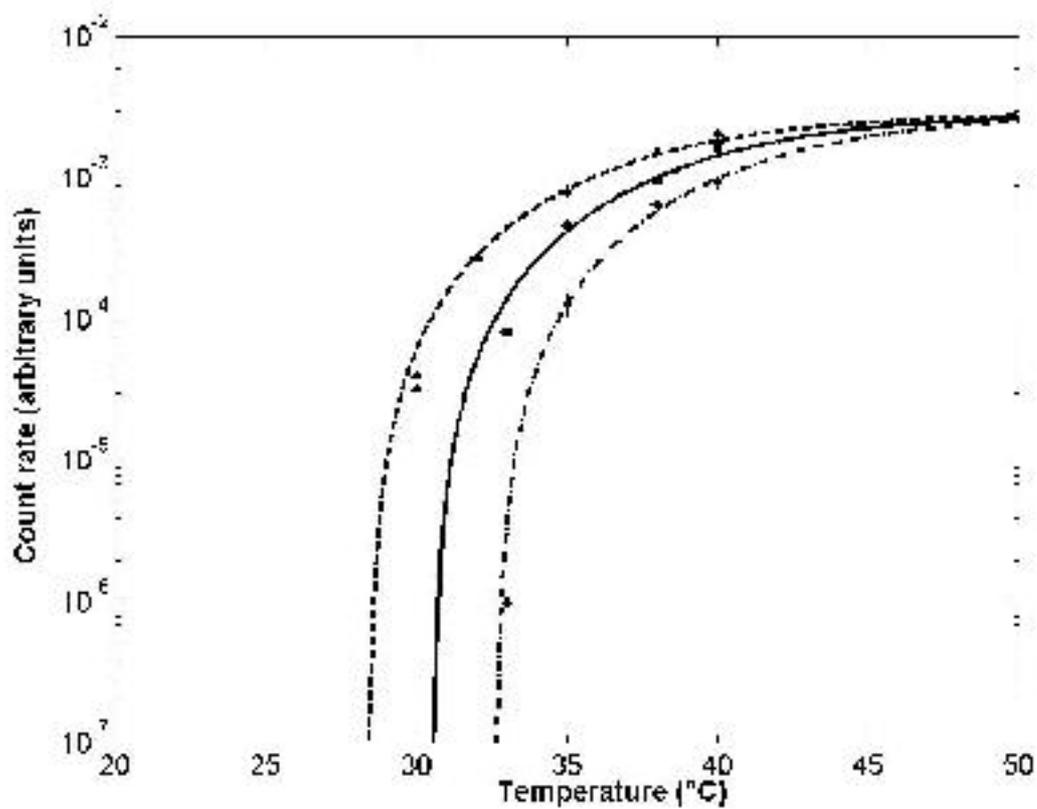

Figure 12

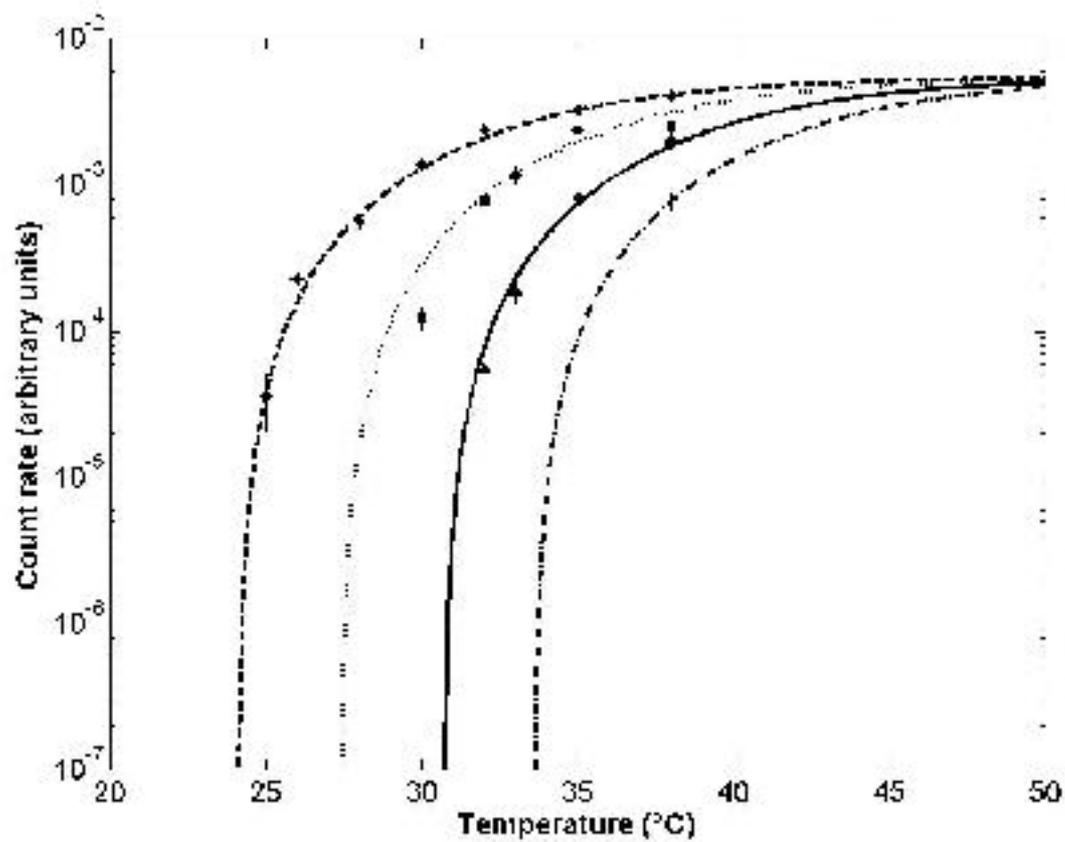

**Figure 13**

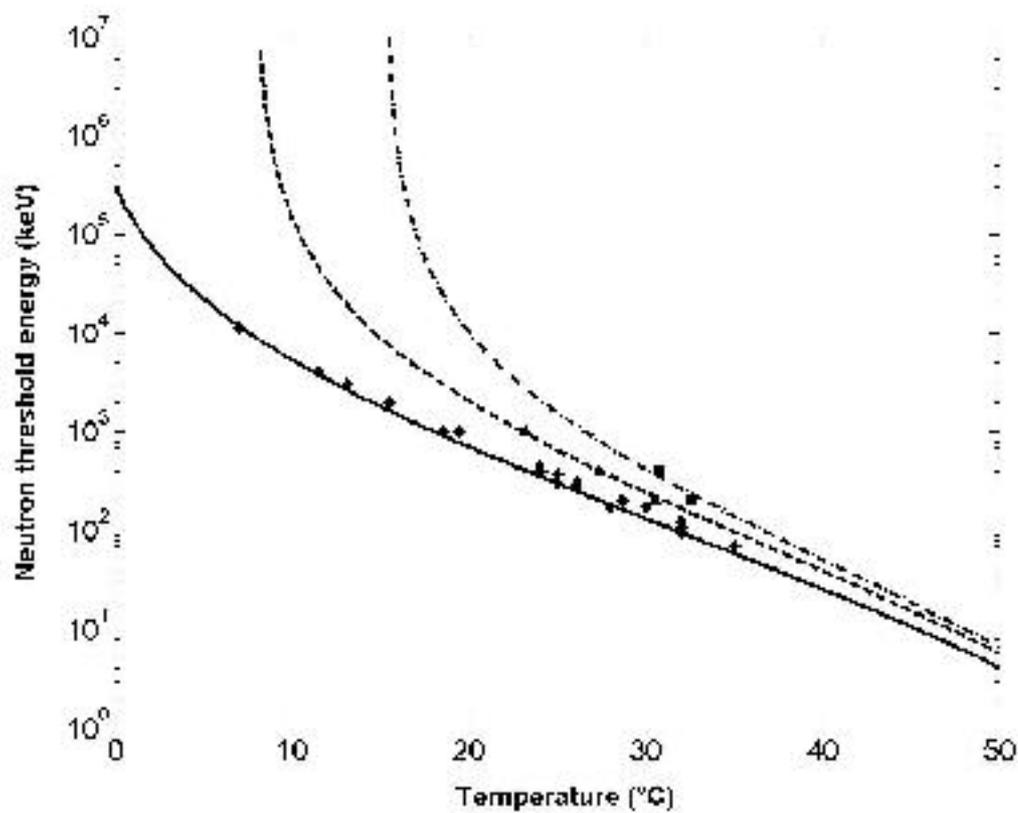

**Figure 14**

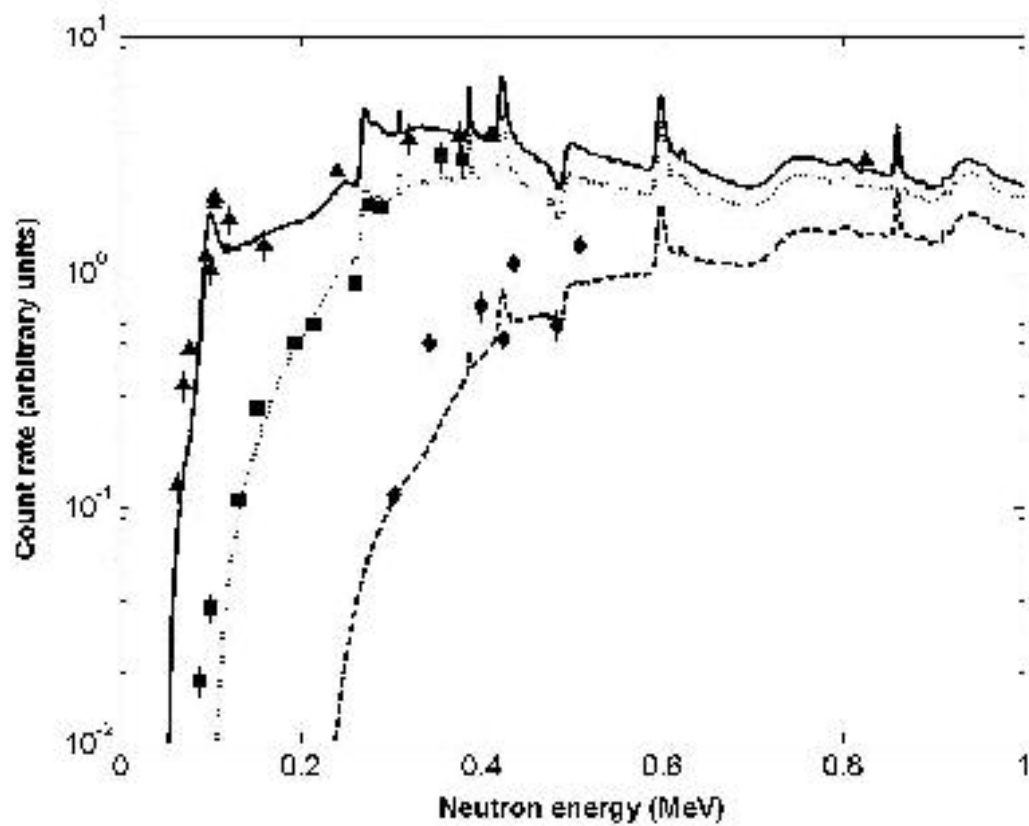

**Figure 15**

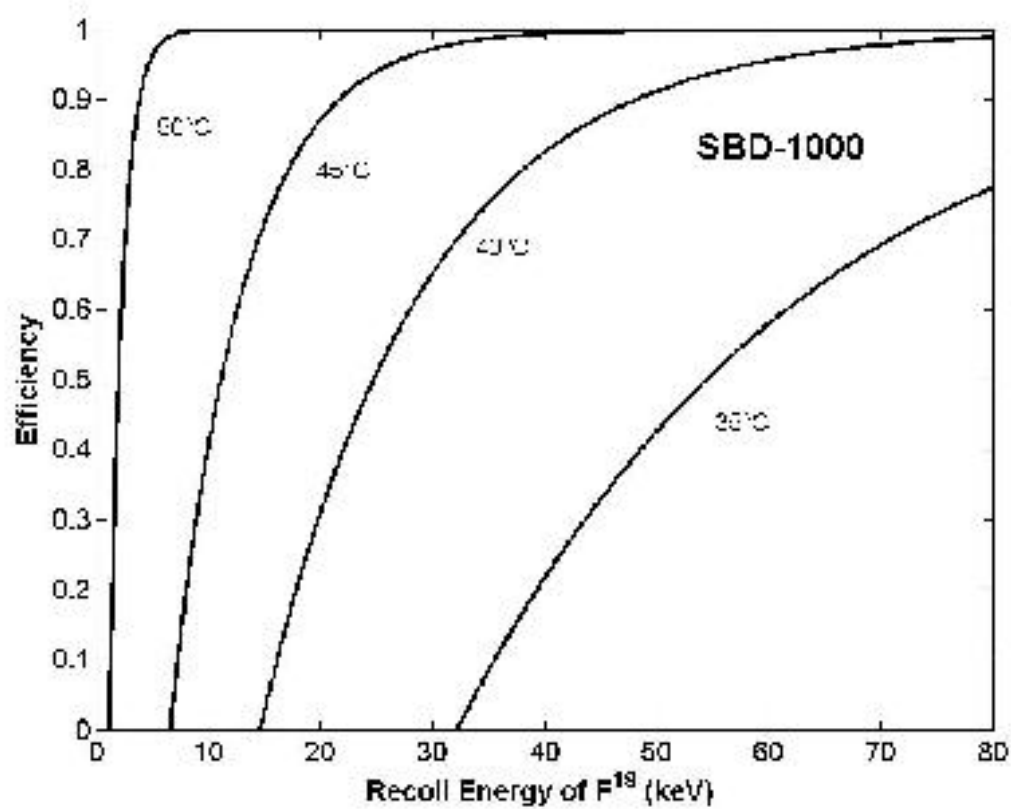

**Figure 16**

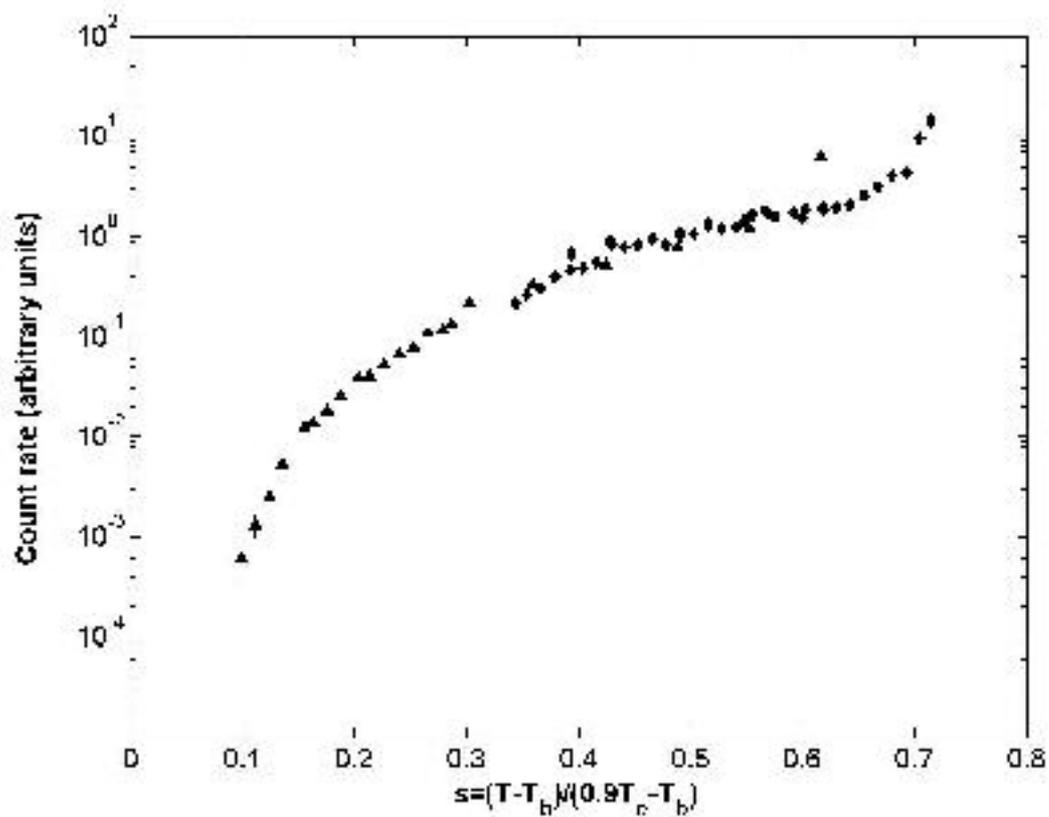

**Figure 17**

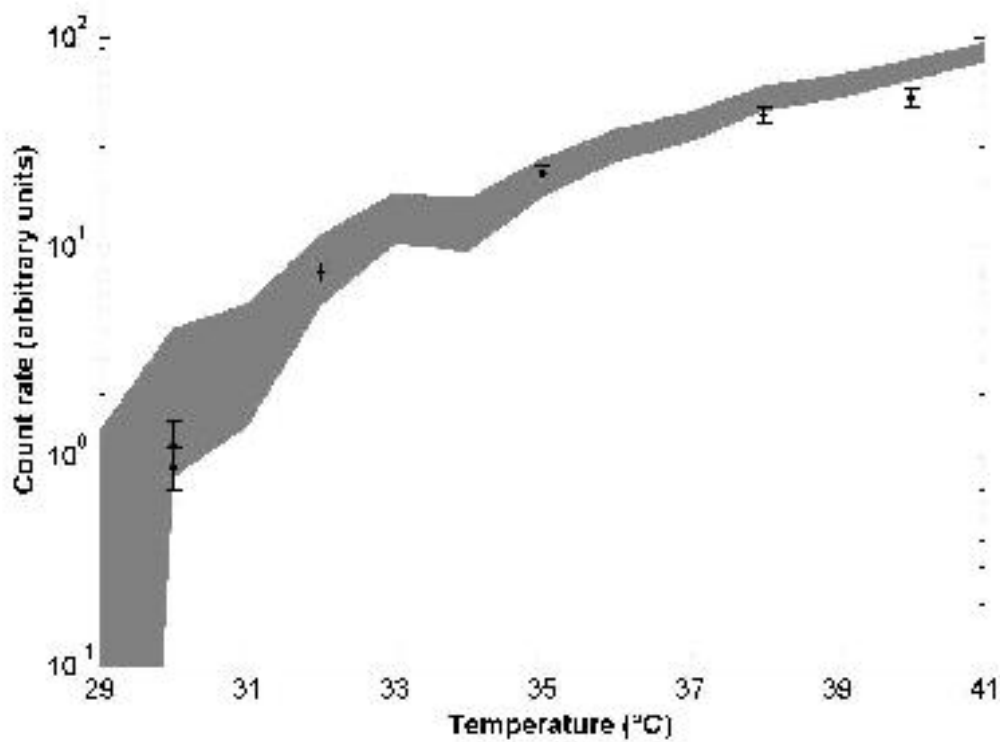

**Figure 18**

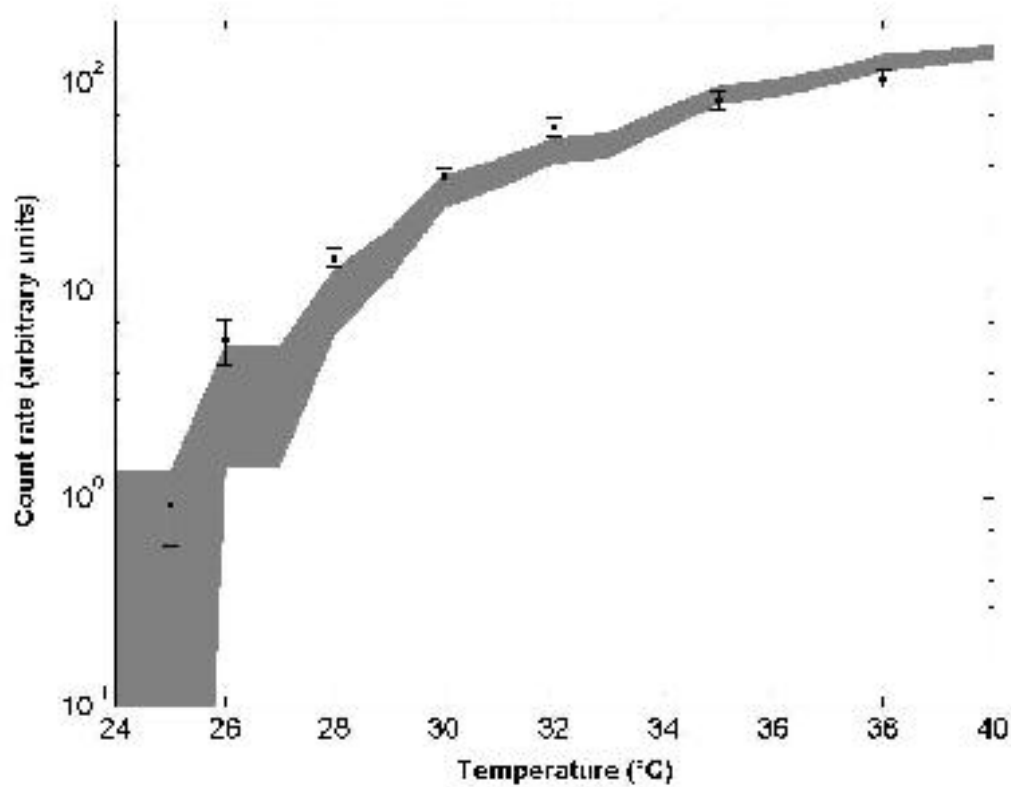

**Figure 19**

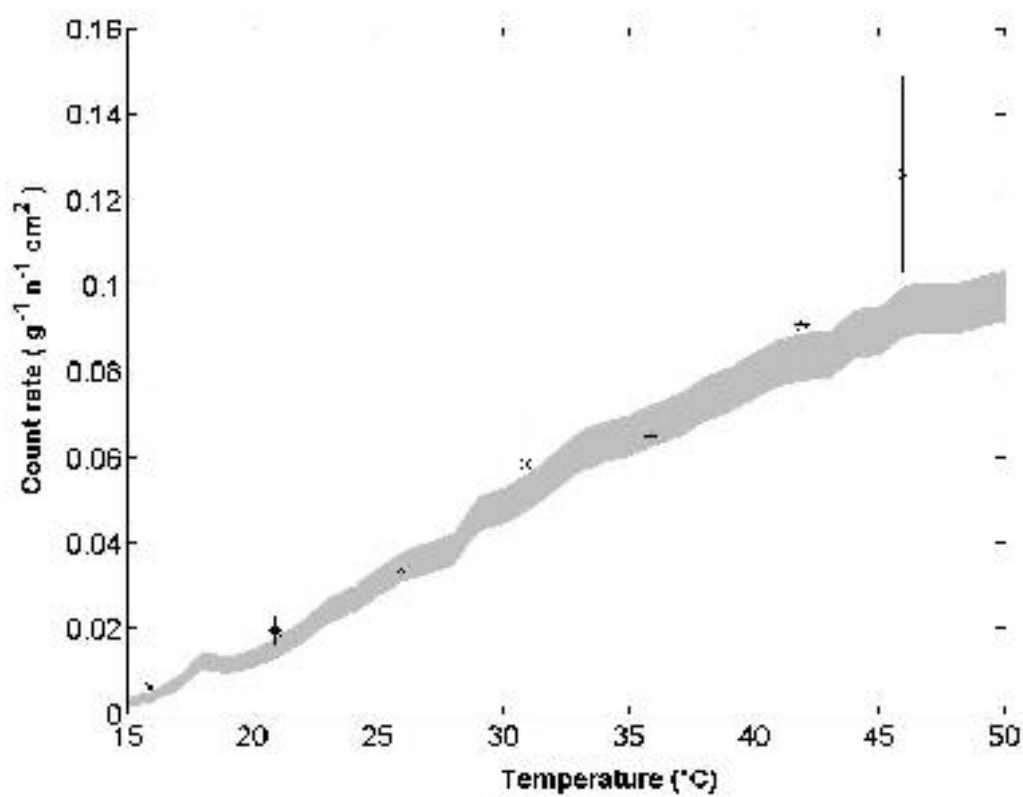

**Figure 20**

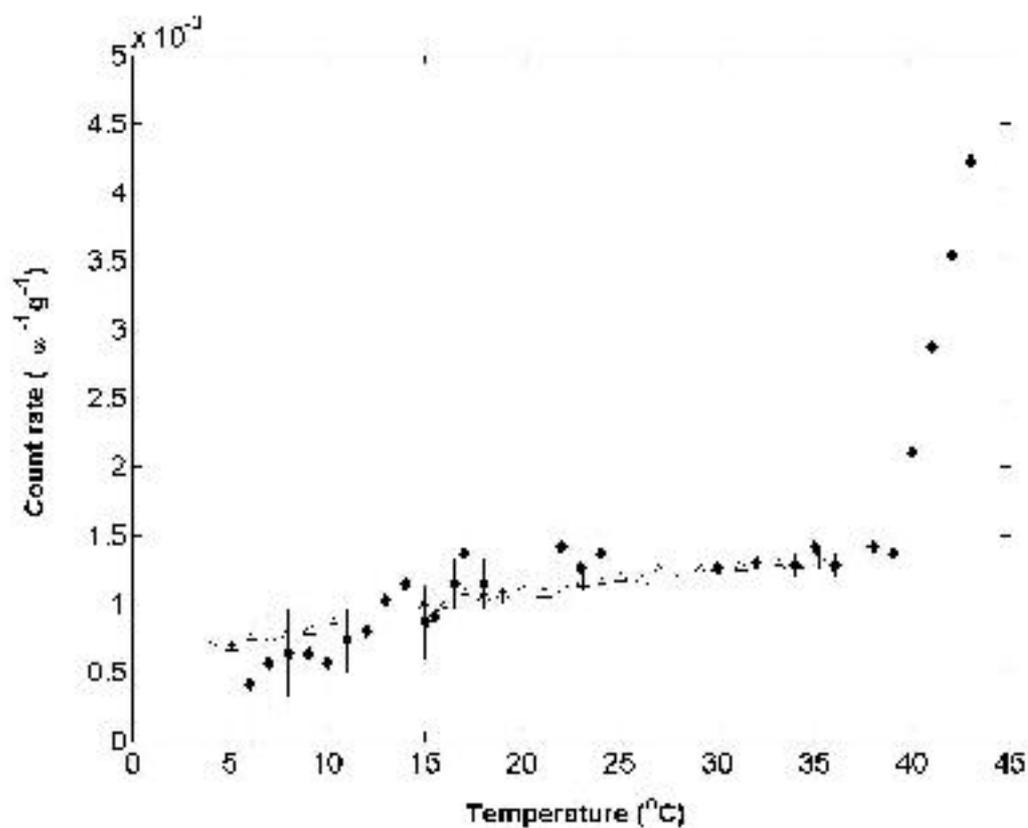

**Figure 21**

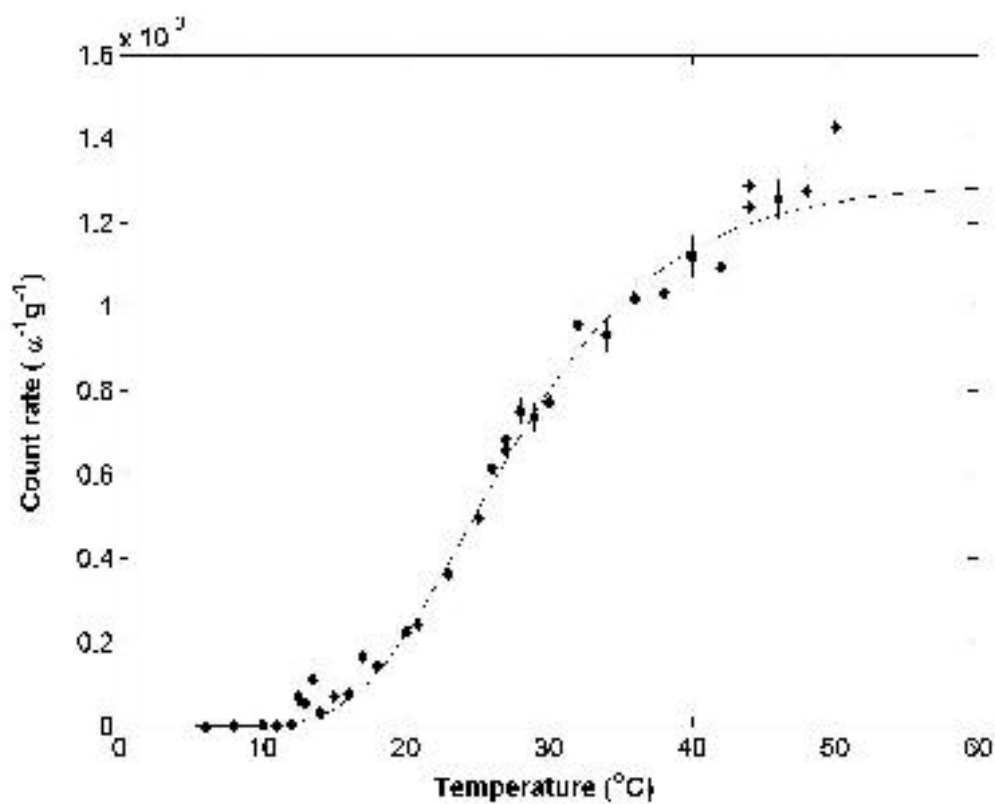

**Figure 22**

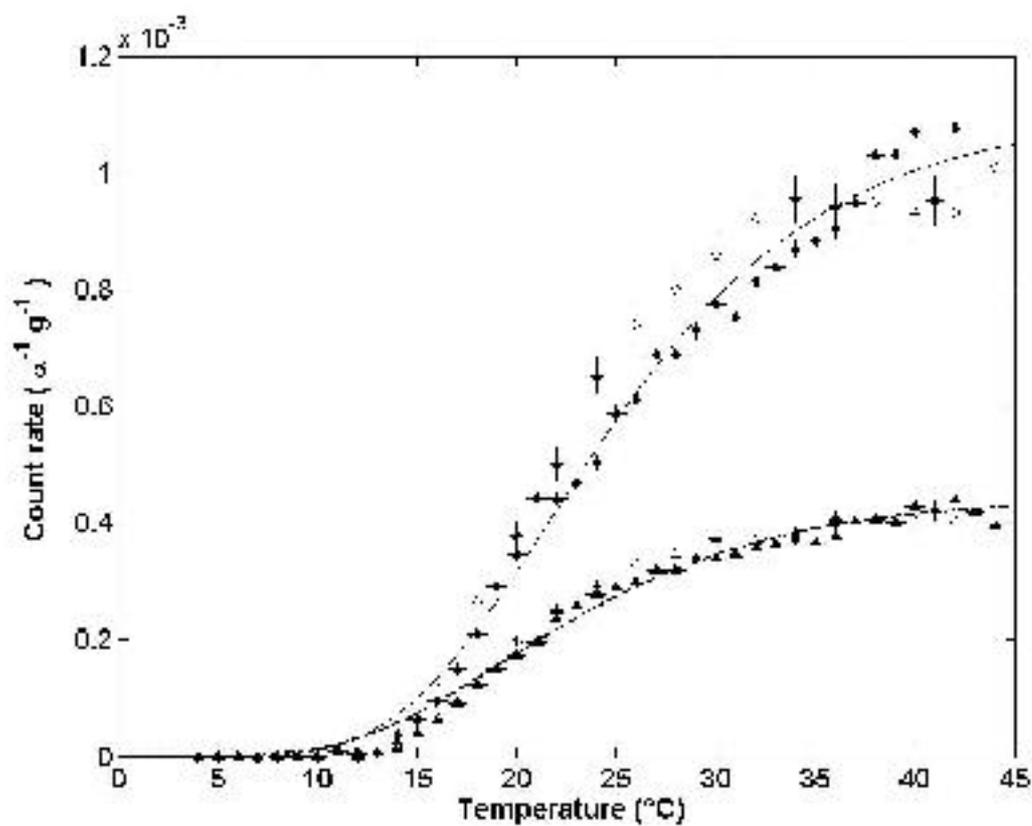

**Figure 23**

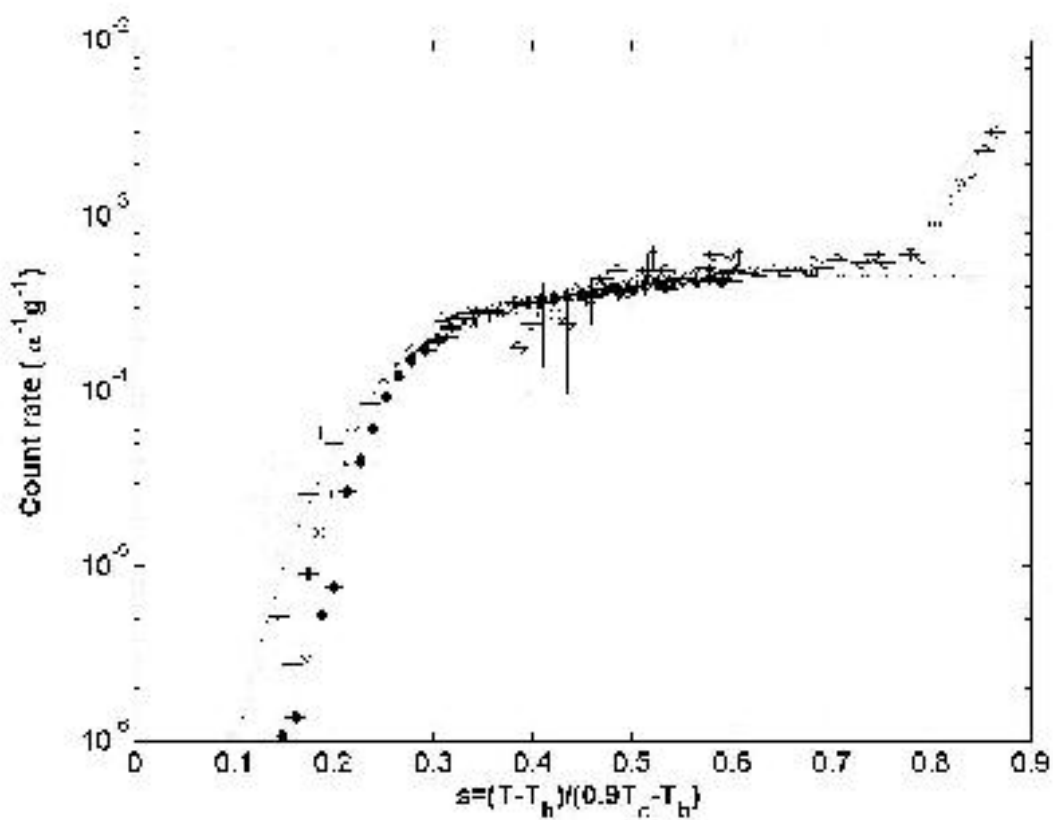

**Figure 24**

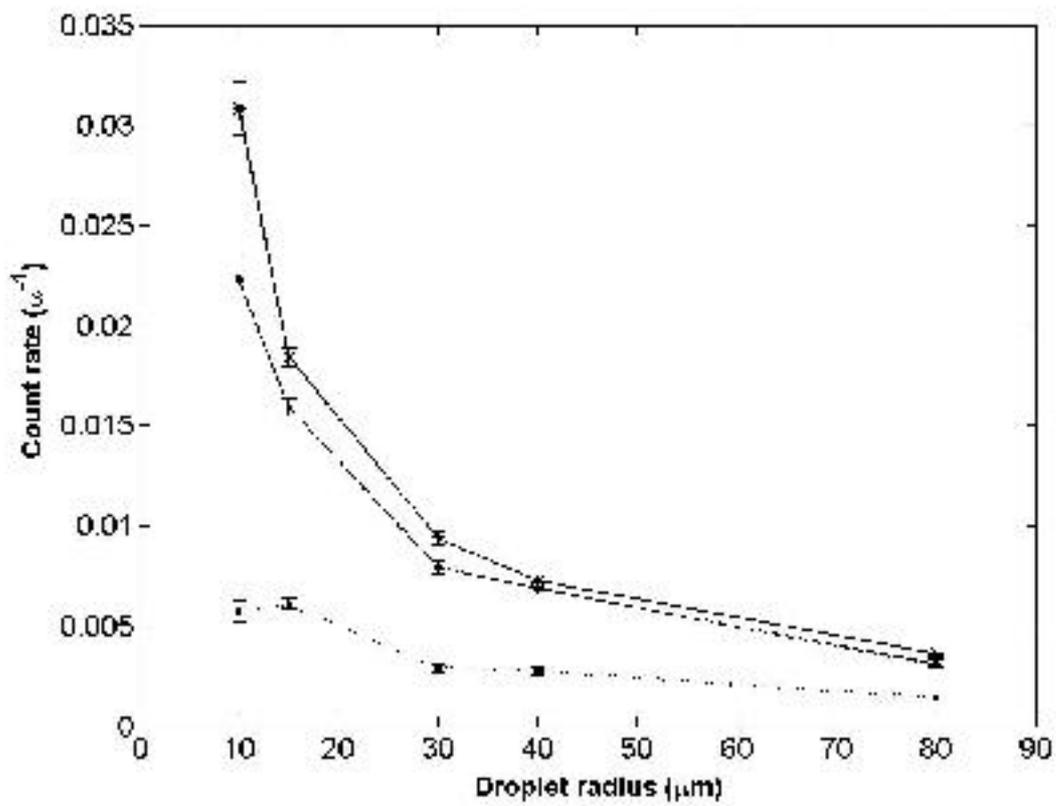

**Figure 25**

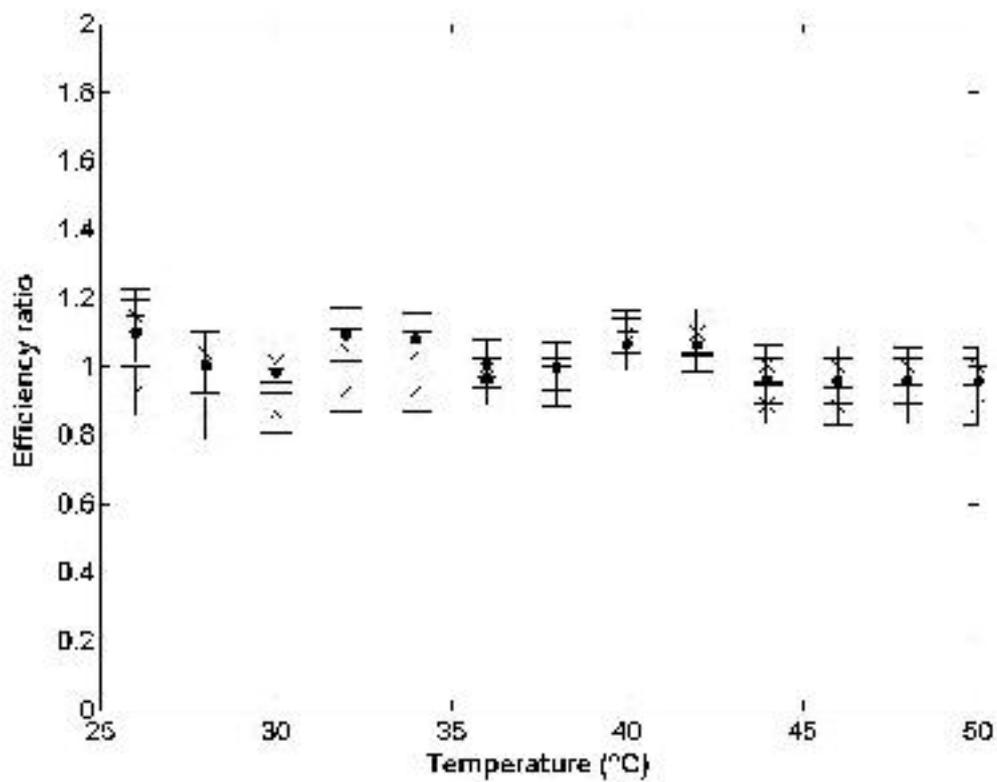

**Figure 26**

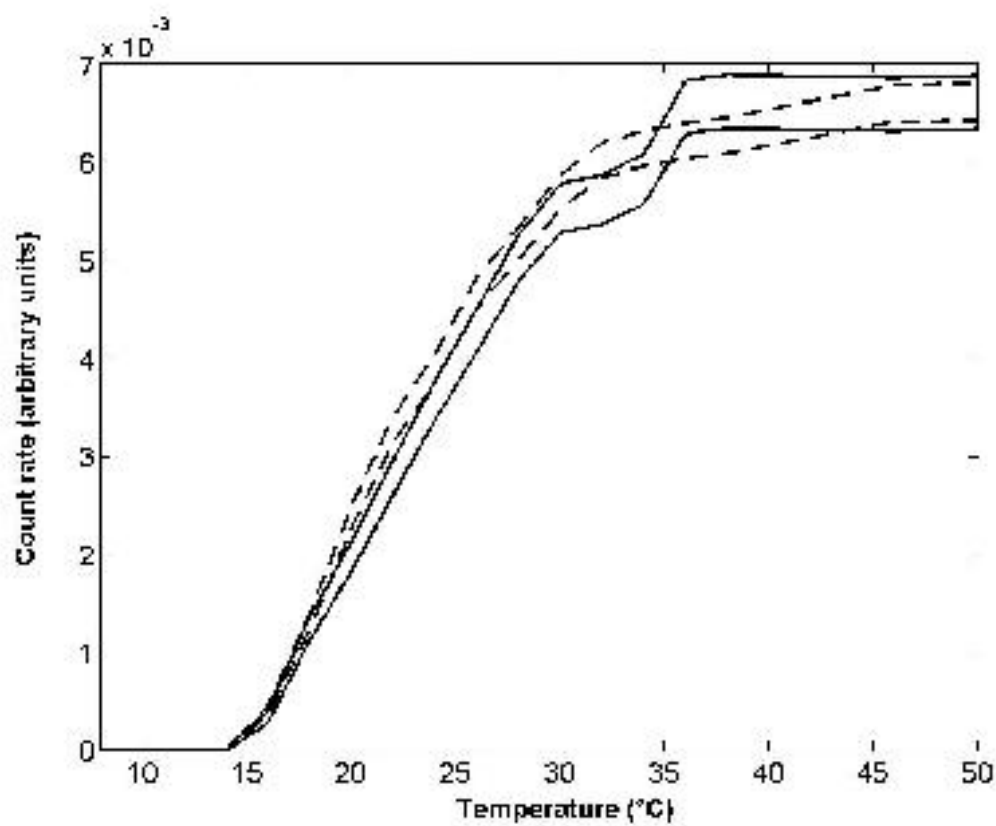

**Figure 27**

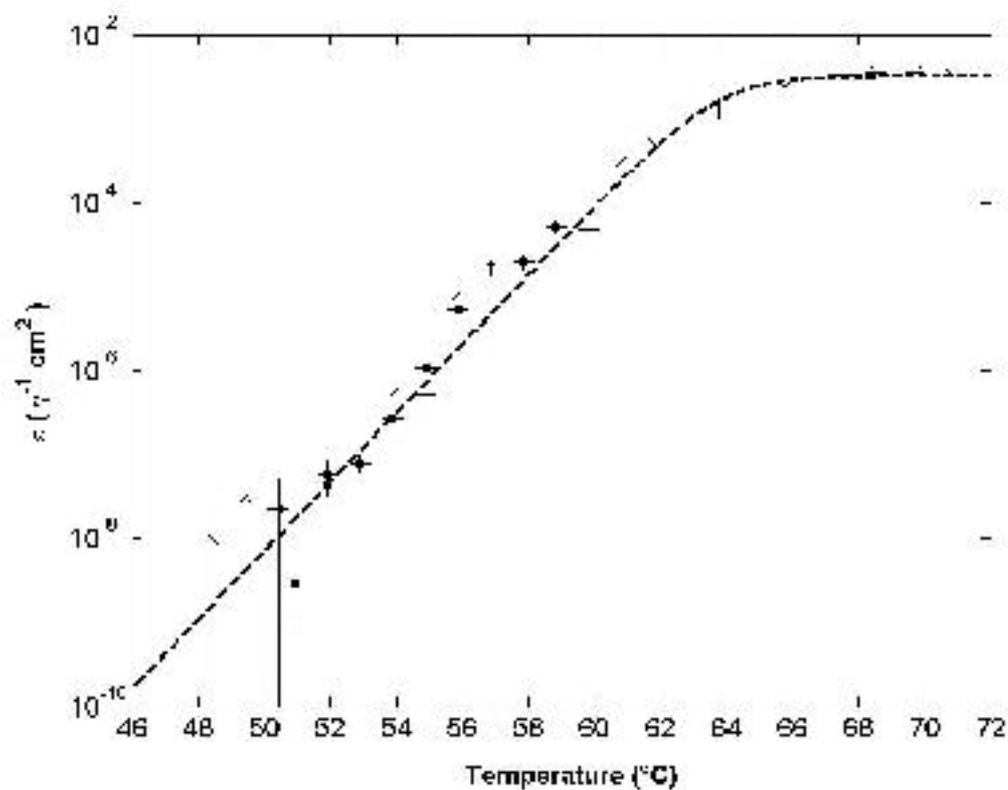

**Figure 28**

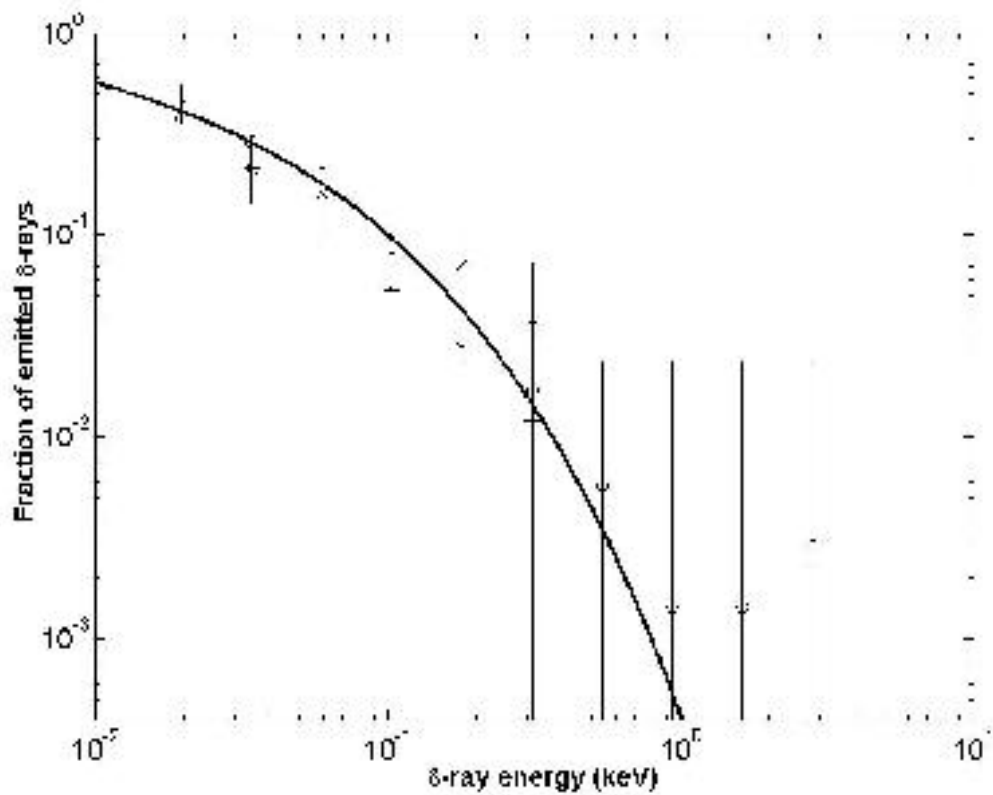

**Figure 29**

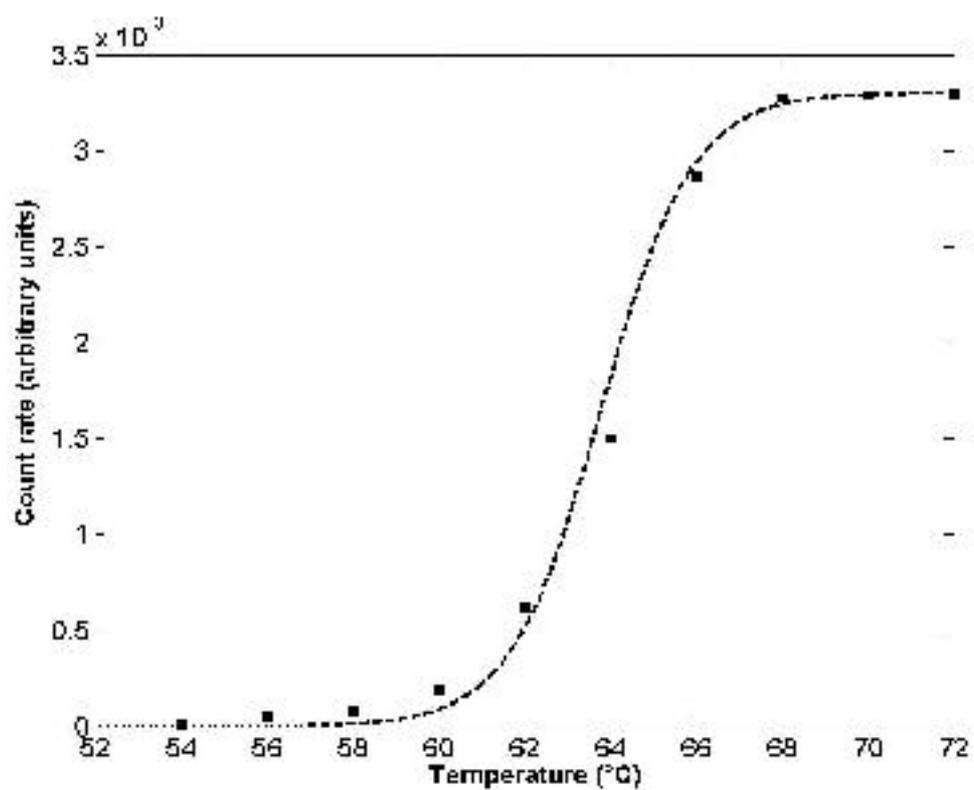

**Figure 30**

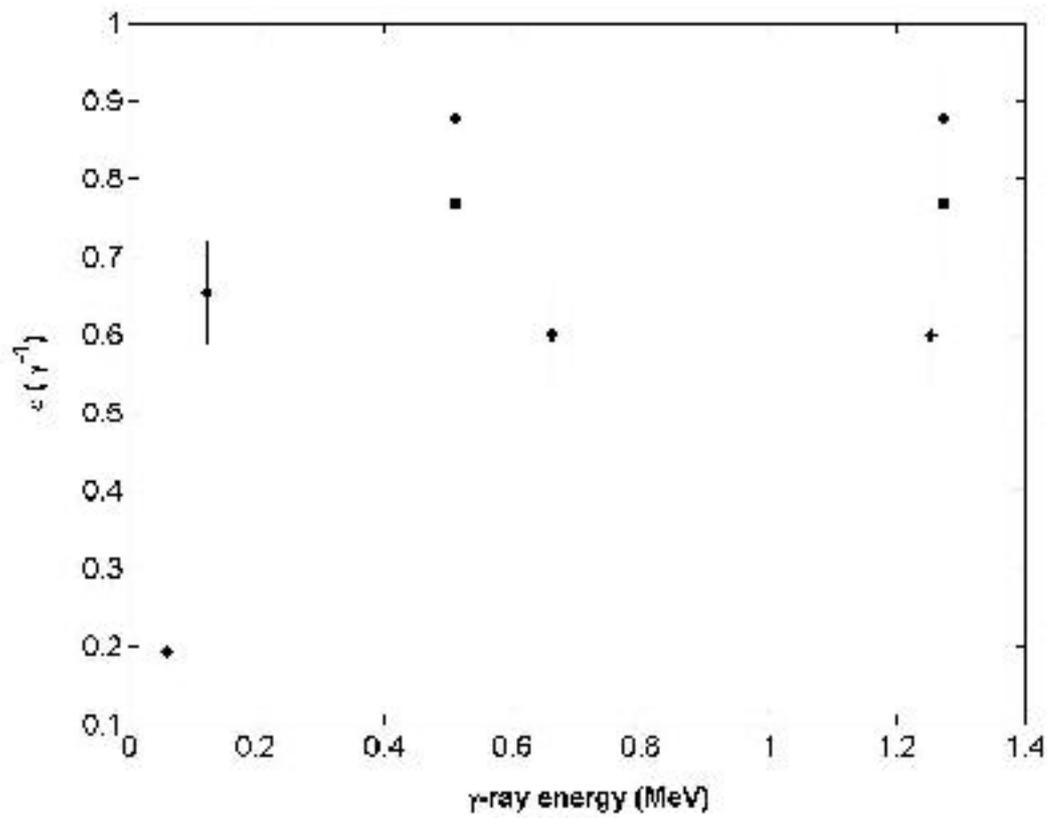

Figure 31

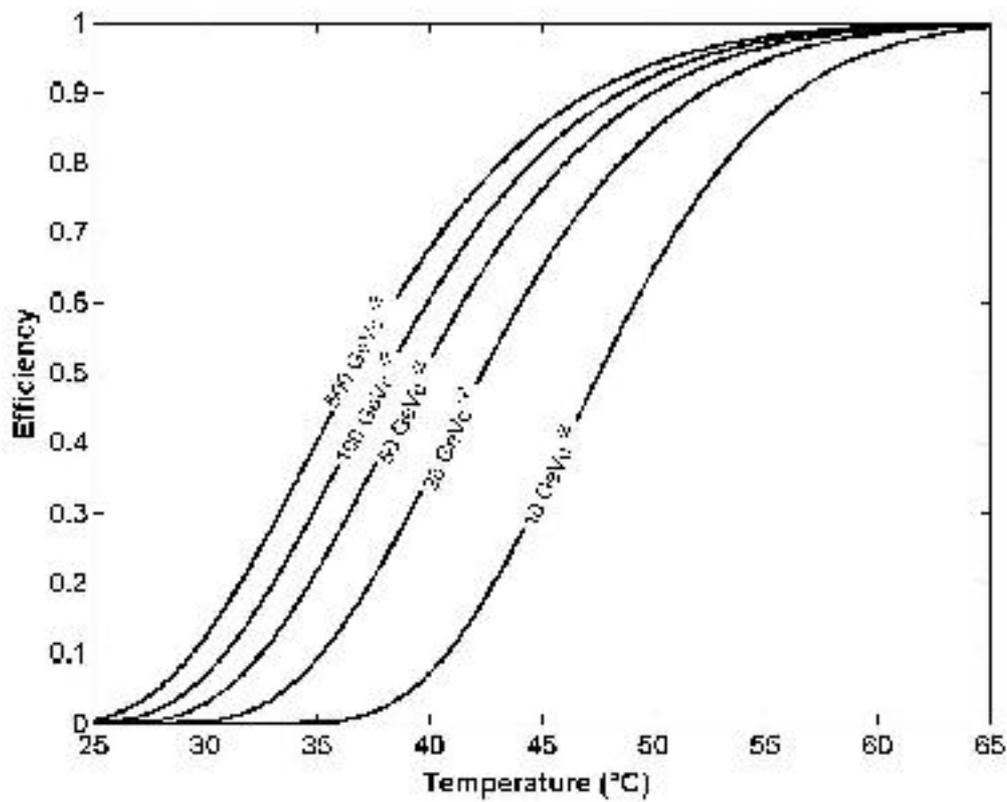

Figure 32

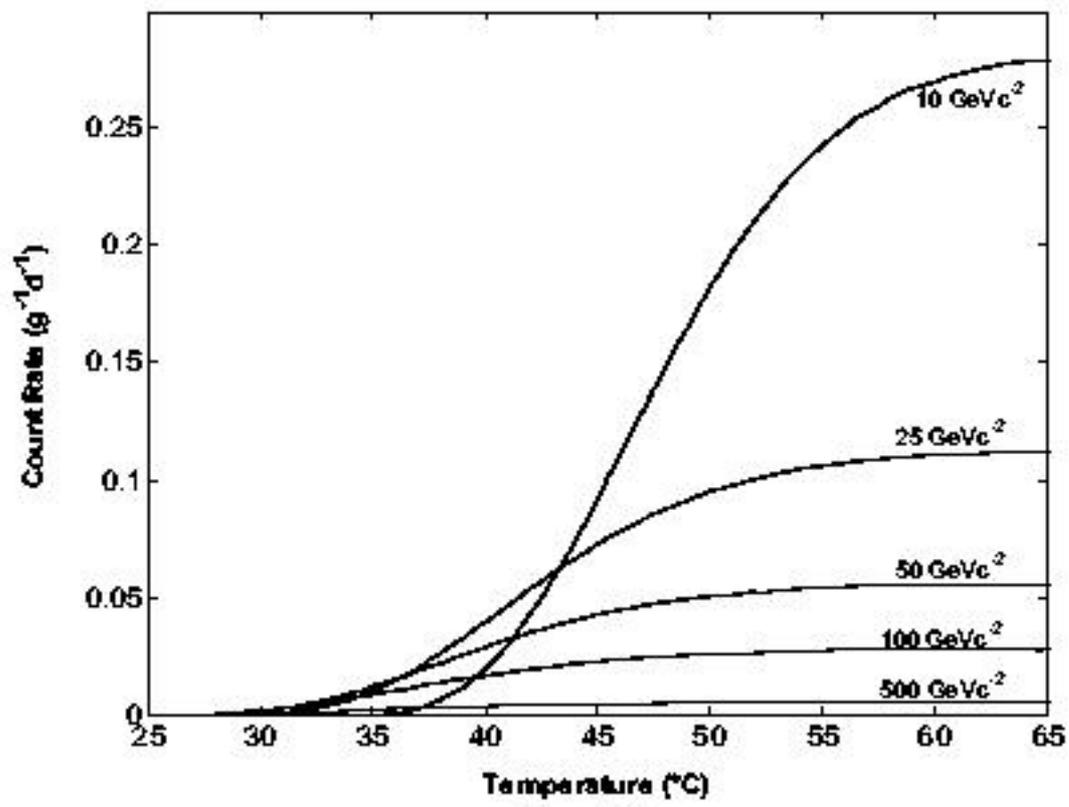

Figure 33